\documentclass[pre,aps,onecolumn,showpacs,showkeys, unsortedaddress, superscriptaddress]{revtex4-2}
\usepackage{amsmath,amssymb,graphicx}
\usepackage{epsfig}
\usepackage{textcomp}
\usepackage[linktocpage,colorlinks=true,allcolors=blue]{hyperref}

\usepackage{orcidlink}

\begin{document}

\title{Large deviations of spatial moments of the Kipnis--Marchioro--Presutti lattice gas following a localized energy release}

\author{Timo Schorlepp\,\orcidlink{0000-0002-9143-8854}}
\email{timo.schorlepp@nyu.edu}
\affiliation{Courant Institute of Mathematical Sciences, New York University, New York, NY 10012, USA}
\author{Baruch Meerson\,\orcidlink{0000-0002-6709-8140}}
\email{meerson@mail.huji.ac.il}
\affiliation{Racah Institute of Physics, Hebrew University of Jerusalem, Jerusalem 91904, Israel}
\author{Pavel Sasorov\,\orcidlink{0000-0001-9257-6565}}
\email{pavel.sasorov@gmail.com}
\affiliation{ELI Beamlines Facility, Extreme Light Infrastructure ERIC, 252 41 Doln\'{\i} B\u{r}e\v{z}any,  Czech Republic}

\begin{abstract}
We study the long-time statistics of spatial moments of the time-dependent energy distribution in a one-dimensional Kipnis--Marchioro--Presutti model following an instantaneous release of energy on a single lattice site. The moments are \(M_n(T)=\sum_i x_i^n u_i(T)\), where \(u_i(T)\) is the energy at site \(i\) at the observation time \(T\). Employing macroscopic fluctuation theory (MFT), we calculate the variance of \(M_n\) for every positive integer \(n\) and determine the non-Gaussian distribution tails for the representative cases \(n=1\) and \(2\). For the first moment, we establish an exact duality, within MFT, to integrated-current statistics for a deterministic flat initial density. For both \(n=1\) and \(n=2\), the large-moment tails are dominated by a single adiabatically evolving doublon: a rapidly propagating pair of narrow solitary pulses of the energy density and of the gradient of the conjugate field. The small-second-moment tail is associated with concentration of the energy near the origin at time \(T\). Comparing single- and two-doublon histories, we predict a dynamical phase transition at a critical value of $M_2$ at which the optimal path breaks reflection symmetry \(x\to -x\). Numerical solutions obtained by direct minimization of the MFT action establish the second-order character of this transition and support our analytical predictions. Finally, we extend calculations of the large- and small-moment tails to arbitrary $n$ and numerically find a transition of the same type for $n=4$.
\end{abstract}

\maketitle

\section{Introduction}

The statistics of sums of many correlated random variables is a central problem in probability theory, statistics, and statistical mechanics~\cite{BulinskiShashkin2007,GiraitisKoulSurgailis2012,MerlevedePeligradUtev2019}. Many macroscopic observables, such as the total energy or magnetization of an interacting system, are sums of local random quantities. Correlations can affect both the scaling and the limiting distribution of these sums, and understanding their rare fluctuations remains a challenging task~\cite{CluselBertin2008}. 

Spectral linear statistics of \(N\times N\) random matrices, \(\sum_{i=1}^{N}f(\lambda_i)\), provide a well-studied example of sums of strongly correlated random variables. Their typical fluctuations have been extensively studied~\cite{Johansson1998,LytovaPastur2009,CundenVivo2014,Forrester2023}, whereas their large-deviation rate functions can exhibit phase transitions in the limit \(N\to\infty\)~\cite{CundenFacchiVivo2016,Valov2024}. Closely related questions arise for linear statistics of particle positions in one-dimensional systems of trapped spinless fermions, including both noninteracting systems and certain interacting models whose ground-state position distributions map onto random-matrix ensembles~\cite{Dean2019,Smith2021fermions}. Recent work has also addressed cumulants and large deviations of linear statistics in the one-dimensional Riesz gas, including its log-gas limit~\cite{LeDoussalSchehr2025}.

Here we consider the more complicated situation of \emph{time-dependent} statistics of correlated sums in an interacting system evolving from a localized initial state. We study spatial moments at an observation time $T$, whose probability distributions depend on both the initial condition and the subsequent dynamics. For diffusive systems with a locally conserved quantity, macroscopic fluctuation theory (MFT)~\cite{Jona} provides a convenient framework for calculating these distributions in the long-time limit. It assigns a probability cost to histories of the density and current fields. To leading exponential order, the probability of a prescribed macroscopic fluctuation is determined by the optimal path: the path that minimizes this cost subject to the initial condition and the constraint on the observable. MFT thus connects the statistics of correlated sums to the dynamics of rare macroscopic fluctuations.

We explore this connection by studying the statistics of spatial moments of a freely evolving Kipnis--Marchioro--Presutti (KMP) lattice gas~\cite{KMP1982}. The KMP model describes stochastic energy exchange between neighboring lattice sites, with each exchange conserving their combined energy. We consider an infinite one-dimensional lattice and assume that a fixed total energy $W$ is released on a single site at time zero. As the energy spreads along the lattice, the site energies $u_i(T)$ become correlated random variables whose sum remains equal to $W$. The spatial moments are the weighted sums
\begin{equation}\label{momentdefinition}
M_n(T)=\sum_i x_i^n u_i(T),\qquad n=1,2,\ldots,
\end{equation}
where $x_i$ is the position of site $i$, measured from the initial release point.  Spatial moments provide a natural characterization of the fluctuating energy packet: $M_1(T)/W$ gives the position of its center of energy, $M_2(T)/W$ measures its mean-square extent relative to the release point, etc. Their probability distributions $\mathcal{P}(M_n)$ therefore quantify fluctuations of transport and spreading that are not captured by the ensemble-averaged energy profile. The initially localized energy release ensures that these moments are finite at every finite observation time $T$, even on the infinite lattice.

At long times, many sites contribute to the spreading energy packet, and a continuum description -- in  terms of fluctuating hydrodynamics \cite{Spohn} leading to MFT -- becomes possible. The MFT approach to the moment statistics builds on earlier studies of integrated-current fluctuations in one-dimensional diffusive lattice gases. By conservation, the current integrated through a fixed point equals the change in the total mass or energy on one side of that point. Integrated current statistics therefore probe a similar observable of the evolving random distribution. Derrida and Gerschenfeld obtained an exact microscopic result for the symmetric simple exclusion process (SSEP) with an annealed step initial density profile \cite{DG2009a}, and formulated the corresponding large-deviation problem for a broader class of diffusive gases within MFT \cite{DG2009b}. Krapivsky and Meerson \cite{KrMe} developed a perturbative MFT approach to typical current fluctuations in this setting and studied larger fluctuations numerically. Subsequent work on the integrated-current fluctuations addressed the extreme-current tails and identified qualitatively different large-deviation mechanisms in different lattice gases \cite{MS2013,MS2014,VMS2014}. 

In the KMP model it was predicted that the optimal path dominating large deviations of the integrated current involves a single doublon: two adjacent localized solitary pulses, one in the energy density and the other in the gradient of the conjugate field, rapidly traveling together in the same direction with slowly evolving amplitudes \cite{MS2013}.

More recently, exact solutions of several nonstationary fluctuation problems have been obtained using the inverse scattering method (ISM), see, e.g., Ref.~\cite{Novikov}. Bettelheim, Smith, and Meerson obtained the full statistics of energy transfer in the KMP model following a localized energy release, first for transfer across the origin \cite{BSM2022a} and then for the energy beyond an arbitrary observation point \cite{BSM2022b}. Mallick, Moriya, and Sasamoto  \cite{MMS2022} solved the MFT problem for the SSEP with an annealed step initial condition, recovering the full current statistics previously obtained by a microscopic method in Ref.~\cite{DG2009a}. Exact equations for the associated spatial correlations have also been obtained by a complementary approach \cite{Grabsch2022}. The ISM was subsequently applied to nonstationary mass transfer in the simple inclusion process with a localized initial condition \cite{BM2024} and to the KMP model with a focus on the energy content in a small spatial interval \cite{BM2024interval}; a short overview of the ISM integrability approach is given in Ref.~\cite{MMS2024}. These developments provide both analytical tools and physical insight for studying the spatial moments of an evolving energy distribution.

Here, employing the MFT, we first calculate the variance of the spatial moments \(M_n\) for all positive integer \(n\). For concreteness, our subsequent analysis of non-Gaussian distribution tails focuses on the first and second moments. As we show, $n = 1$ and $n = 2$ are representative of odd and even $n$, respectively.
For the first moment, we establish an exact duality, within MFT, to integrated-current statistics for a deterministic flat initial density. The large-\(|m_1|\) and large-\(m_2\) tails, where \(m_n=M_n/(WT^{n/2})\), are described by a single adiabatically evolving rapidly traveling doublon, which transports energy far from the origin.  Comparing single- and two-doublon histories in the large-\(m_2\) regime, together with the symmetric optimal path near the expected moment, we predict a dynamical phase transition at an intermediate value \(m_2^*\), beyond which the optimal path breaks reflection symmetry. The
small-\(m_2\) tail  is associated with concentration of the energy near the origin at the observation time $T$. 

We test our analytical predictions and characterize the symmetry-breaking transition numerically by direct minimization of the MFT action. The numerical results establish that the transition is of second order and determine its critical point. 
The back-and-forth iteration algorithm of Chernykh and Stepanov \cite{CS2001}, although successfully used in many instanton computations,  encounters convergence difficulties in the KMP regimes studied here, especially for the large-\(m_2\) and small-\(m_2\) tails. (These difficulties have also left the large-current asymptotics of Ref.~\cite{MS2013} without numerical verification for more than a decade.) Here we avoid these difficulties by treating the minimization as an optimal-control problem~\cite{GVE2019,SGMG2022,STGS2023,SSM2023,BSG2024,DeGiuli25,SG2026}. Imposing the moment constraint directly allows us to compute the nonconvex part of the rate function on the symmetry-broken branch. Direct action minimization has previously been used for boundary-driven lattice gases~\cite{BKP2012}. By virtue of the exact duality for $n=1$, our computations also provide the first numerical solutions of the KMP MFT equations for large integrated currents. 

The remainder of the paper is structured as follows. Section~\ref{sec:MFT} formulates the spatial-moment problem within MFT, and Sec.~\ref{sec:integrals} presents the integrals of motion. Section~\ref{sec:Gaussian} derives the Gaussian statistics and the moment variances for arbitrary positive integer $n$. Section~\ref{sec:n1} establishes the duality between first-moment and integrated-current statistics and obtains the large-$|m_1|$ tail. Section~\ref{sec:m2} addresses the large- and small-$m_2$ tails and presents the argument for a symmetry-breaking dynamical phase transition. The numerical method is described in Sec.~\ref{sec:numericalalgorithm}, followed by comparisons with theory for $n=1$ and $n=2$ in Secs.~\ref{sec:numn1} and~\ref{sec:numn2}, including the characterization of the transition. Section~\ref{sec:higher} extends the analysis to $n\geq3$. Section~\ref{sec:discussion} summarizes and discusses our results. Appendix~\ref{app:currentcorrection} explains the $O(1)$ prefactor correction in the extreme-current asymptotics used for $n=1$. Appendices~\ref{app:m2doublon} and~\ref{app:smallm2} provide supporting calculations for the large- and small-$m_2$ tails, respectively.

\section{Fluctuating Hydrodynamics and MFT}\label{sec:MFT}

The starting point of the MFT \cite{Jona} is fluctuating hydrodynamics  \cite{Spohn}: a stochastic PDE for the coarse-grained
energy density (which we simply call density) $u(x,t)$:
\begin{equation}\label{Langevin}
\partial_t u = \partial_{x} \left[D(u)\partial_x u +\sqrt{\sigma(u)}\,\xi(x,t)\right],
\end{equation}
where $\xi(x,t)$ is space-time white Gaussian noise
\begin{equation}\label{noisecorrelation}
\langle \xi(x_1,t_1) \xi(x_2,t_2) \rangle = \delta(x_1-x_2) \delta(t_1-t_2)\,.
\end{equation}
For the KMP model the transport coefficients are $D(u) = 1$ and $\sigma(u) = 2 u^2$.
The MFT equations are obtained via the saddle-point approximation to the path integral for Eq.~(\ref{Langevin}).
We define $v=-\partial_x p$, where $p(x,t)$ is the conjugate field. The resulting MFT equations are \cite{Jona,Tailleur,DG2009b,KrMe,MS2013,BSM2022a}
\begin{eqnarray}
  \partial_t u &=& \partial_{x}(\partial_x u+2u^2 v), \label{qpa} \\
  \partial_t v &=&  \partial_{x} (-\partial_x v+ 2u v^2). \label{qpb}
\end{eqnarray}
These equations are Hamiltonian. The Hamiltonian,
\begin{equation}\label{H}
H[u(x,t),v(x,t)] = \int_{-\infty}^{\infty} dx\,\left(v \partial_x u +u^2 v^2\right) 
\end{equation}
is conserved in time. 
We suppose that the whole energy $W$ is released  at $t=0$ on a single lattice site:
\begin{equation}\label{incond0}
u(x,t=0) = W\delta(x)\,.
\end{equation}
Obviously, the energy is conserved under the dynamics of Eqs.~(\ref{qpa}) and (\ref{qpb}):
\begin{equation}\label{energy}
\int_{-\infty}^{\infty} u(x,t) dx = W.
\end{equation}
We condition the process on the value of the moment $M_n$ ($n=1,2,\dots$) at the observation time $T$:
\begin{equation}\label{constraint0}
\int_{-\infty}^{\infty} x^n\,u(x,T)\,dx = M_n\,.
\end{equation}
This constraint is imposed with the help of a Lagrange multiplier $\Lambda$. Minimizing the constrained action with respect to both the dynamics and the \textit{a priori} unknown final density $u(x,T)$, we obtain the following boundary condition at $t=T$:
\begin{equation}\label{fincond0}
v(x,T) =-\Lambda nx^{n-1}\,,
\end{equation}
where $\Lambda$ is to be ultimately set by Eq.~(\ref{constraint0}). 
Once the optimal path is determined, the action $S\simeq -\ln \mathcal{P}$ is given by
\begin{equation}\label{action0}
    S=\int_0^T dt \int_{-\infty}^{\infty} dx\, u^2(x,t) v^2(x,t).
\end{equation}
Alternatively, when $M_n(\Lambda)$ is determined, one can find $S$ from the Legendre relation $dS/dM_n = \Lambda$.
Upon rescaling $x/\sqrt{T} \to x$, $t/T \to t$, $u\sqrt{T}/W\to u$ and $W v\to v$, we bring the equations to a rescaled form.
The rescaled Eqs.~(\ref{qpa}) and (\ref{qpb}) do not change.
The rescaled two-time boundary conditions are
\begin{equation}\label{incond1}
u(x,0) = \delta(x)
\end{equation}
and
\begin{equation}\label{fincond1}
v(x,1) =-n\lambda x^{n-1}\,,
\end{equation}
where $\lambda=WT^{(n-1)/2}\Lambda$ is the rescaled Lagrange multiplier, to be determined by the rescaled constraint
\begin{equation}\label{constraint1}
\int_{-\infty}^{\infty} x^n\,u(x,1)\,dx = m_n\equiv \frac{M_n}{W T^{\frac{n}{2}}}\,.
\end{equation}
The action (\ref{action0}) becomes
\begin{equation}\label{SRW}
S= \sqrt{T} s(\Delta m_n)\,,
\end{equation}
where $\Delta m_n=m_n-\bar{m}_n$, $\bar{m}_n$ is the expected value of the rescaled moment $m_n$ at $t=1$, and the rescaled action
\begin{equation}\label{action}
    s(\Delta m_n)=\int_0^1 dt \int_{-\infty}^{\infty} dx\, u^2 v^2 
\end{equation}
plays the role of the rate function.
Alternatively, having found $m_n(\lambda)$, one can find $s(m_n)$ from the  relations
\begin{equation}\label{shortcut}
  \frac{ds}{dm_n} = \lambda\,,\quad \text{and} \quad s(\Delta m_n=0)=0\,.
\end{equation}
The last equality reflects the fact that the action vanishes at the expected value of $m_n$. The latter can be obtained by a straightforward calculation: 
\begin{equation}\label{barmn}
\bar{m}_n =
\begin{cases}
0, &n=2k+1,\;k=0,1,2,\dots\\
\frac{4^k \Gamma \left(k+\frac{1}{2}\right)}{\sqrt{\pi }}, & n=2k,\; k=1,2,\dots \,,\\
\end{cases}
\end{equation}
where $\Gamma(\dots)$ is the gamma function.

\section{Integrals of motion}\label{sec:integrals}

We have already mentioned the obvious integral of motion of the MFT equations -- Eq.~(\ref{energy}), which describes the energy conservation.  The similar integral of $v$,  $\int_{-\infty}^{\infty}v\,dx$, is not directly usable because, by virtue of the terminal boundary condition (\ref{fincond1}), $v$ does not decay at $x\to \infty$. We note, however, that at sufficiently large $|x|$, the field $v(x,t)$ solves the backward heat equation with the initial condition given by the terminal boundary condition~(\ref{fincond1}). As one can check, this background  solution is polynomial:
\begin{equation}\label{vbackground}
v_{\mathrm{bg}}(x,t)=-\lambda n!
\sum_{k=0}^{\lfloor(n-1)/2\rfloor}
\frac{(1-t)^k x^{n-1-2k}}{k!(n-1-2k)!}\,,
\end{equation}
where $\lfloor(n-1)/2\rfloor$ denotes 
the largest integer not exceeding  $(n-1)/2$. The general expression (\ref{vbackground}) will be useful for the linear theory in Sec.~\ref{sec:Gaussian}. In particular, for $n=1$ and $2$ Eq.~(\ref{vbackground}) gives $-\lambda$ and $-2\lambda x$, respectively.  Assuming that $v-v_{\mathrm{bg}}$ is integrable, and that the associated boundary terms vanish, Eq.~(\ref{qpb}) yields
\begin{equation}\label{vregularizedconservation}
\int_{-\infty}^{\infty}(v-v_{\mathrm{bg}})\,dx=\text{const} = 0. 
\end{equation}
Indeed, since this regularized integral vanishes at $t=1$, it therefore vanishes at all times.

The two integrals of motion mentioned above are common for a broad family of diffusive lattice gases. The KMP model is special because here there is also an \emph{infinite} number of additional, nontrivial integrals of motion, which originate from the fact that Eqs.~(\ref{qpa}) and (\ref{qpb}) are in principle (although not necessarily in practice) completely integrable by the ISM \cite{BSM2022a}. The simplest of these integrals is
\begin{equation}\label{uv}
I_0=\int_{-\infty}^{\infty} u v\, dx=\text{const}\,.
\end{equation}
Here we present three more integrals of motion; the list could be continued:
\begin{eqnarray}
  I_1 &=&  \int_{-\infty}^{\infty} (vu_x+u^2v^2) dx  = \int_{-\infty}^{\infty} (-u v_x+u^2v^2) dx=\text{const}, \label{I1}\\
  I_2&=& \int_{-\infty}^{\infty} (vu_{xx}+4uu_xv^2+u^2vv_x+2u^3v^3) dx =\text{const},\label{I2}\\
  I_3&=& \int_{-\infty}^{\infty}(v u_{xxx}+6u u_{xx}v^2+5u_x^2v^2+6u u_x v v_x+u^2v v_{xx}+16u^2u_xv^3+6u^3v^2v_x+5u^4v^4)\,dx =\text{const}. \label{I3}
\end{eqnarray}
The integral $I_1$ is the Hamiltonian of the MFT equations, Eq.~(\ref{H}), so its conservation is obvious.
These conservation laws hold for $0<t\leq1$. We exclude the singular initial endpoint $t=0$, where $u(x,0)=\delta(x)$. The integrals converge even though $v$ grows polynomially at large $|x|$ for $n\geq2$: every term contains a factor of $u$ or of its derivatives, which decay faster than any power of $x$ for $t>0$.

For $n=1$ we have $I_0=-\lambda$, $I_1=\lambda^2\int_{-\infty}^{\infty} u(x,1)^2 dx$  and
$I_2=-2\lambda^3 \int_{-\infty}^{\infty} u(x,1)^3 dx$. The values of the higher-order integrals involve the $x$-derivatives of $u(x,1)$. 

For $n=2$ we obtain $I_0=-2\lambda \int_{-\infty}^{\infty} x u(x,1) dx= -2\lambda m_1$. Further, $I_1=2\lambda +4\lambda^2 \int_{-\infty}^{\infty}x^2 u(x,1)^2 dx$, and
$I_2 = -12 \lambda^2 \int_{-\infty}^{\infty}x u(x,1)^2 dx-16 \lambda^3 \int_{-\infty}^{\infty}x^3 u(x,1)^3 dx$, etc. For mirror-symmetric solutions with respect to $x=0$, with $u$ even and $v$ odd in $x$, the even-order integrals $I_0$, $I_2$, $I_4$, $\ldots$ vanish: in every term of $I_k$ the number of factors $v$ plus the number of $x$-derivatives equals $k+1$, so that the integrand is odd for even $k$.

\section{Typical (Gaussian) fluctuations for arbitrary $n$}\label{sec:Gaussian}
To leading order at long times, typical fluctuations of $M_n$ are Gaussian. They correspond to small $\lambda$ and can be analyzed via linearization of Eqs.~(\ref{qpa}) and (\ref{qpb})
around the  expected (deterministic) solution 
\begin{equation}\label{rhobar}
\bar{u}(x,t) = \frac{1}{\sqrt{4\pi t}}\,e^{-\frac{x^2}{4t}}\,,\quad v(x,t) \equiv 0\,,
\end{equation}
see Ref.~\cite{KrMe}. The linearized equations for the small density correction $\delta u(x,t)=O(\lambda)$ and $v(x,t)=O(\lambda)$ are the following:
\begin{eqnarray}
  \partial_t \delta u &=& \partial_{x}^2 \delta u+2\partial_x \left(\bar{u}^2 v\right), \label{qpalin} \\
  \partial_t v &=& - \partial_{x}^2 v. \label{qpblin}
\end{eqnarray}
The boundary conditions in time are $\delta u(x,t=0)=0$ and Eq.~(\ref{fincond1}).

\subsection{$n=1$}\label{sec:Gn1}

We start with $n=1$. Solving Eq.~(\ref{qpblin}) backward in time with the ``initial'' condition~(\ref{fincond1}), we obtain simply $v(x,t) = v(x,t=1) =-\lambda =\text{const}$: the solution coincides with the ``initial'' condition.  Therefore,
the action as a function of $\lambda$ is [see Eq.~(\ref{action})]
\begin{equation}\label{slambdan=1}
s_1(\lambda) \simeq \lambda^2 \int_0^1 dt\int_{-\infty}^{\infty} \bar{u}(x,t)^2 dx = \frac{\lambda^2}{\sqrt{2\pi}}\,.
\end{equation}
Using Eq.~(\ref{shortcut}), we obtain
\begin{equation}\label{s1linm}
s_1(m_1) \simeq\sqrt{\frac{\pi}{8}}\,m_1^2\,.
\end{equation}
The small correction to the deterministic solution for the density follows from Eq.~(\ref{qpalin}) in a straightforward manner, and we obtain
\begin{equation}\label{deltau1}
\delta u (x,t) = \frac{\lambda e^{-\frac{x^2}{4t}}\text{erf}\,\left(\frac{x}{\sqrt{4t}}\right)}{\sqrt{4\pi t}}\,.
\end{equation}

\subsection{$n=2$}\label{sec:Gn2}

Here too the solution of Eq.~(\ref{qpblin}) coincides with the ``initial'' condition: $v(x,t) = v(x,t=1)= -2\lambda x$.
Therefore,
the action as a function of $\lambda$ is 
\begin{equation}\label{slambdan=2}
s_2(\lambda) \simeq  \int_0^1 dt \int_{-\infty}^{\infty} \bar{u}(x,t)^2  (-2\lambda x)^2 dx = \frac{2}{3} \sqrt{\frac{2}{\pi }} \lambda ^2\,.
\end{equation}
With Eq.~(\ref{shortcut}), this yields
\begin{equation}\label{s2linm}
s_2(m_2) \simeq\frac{3}{8} \sqrt{\frac{\pi }{2}} \left(m_2-\bar{m}_2\right)^2\,,
\end{equation}
where 
\begin{equation}\label{barm2}
\bar{m}_2 = \int_{-\infty}^{\infty} x^2 \bar{u}(x,t=1) \,dx =2
\end{equation}
is the expected (mean-field) value of the  rescaled second moment.

\subsection{General $n$}
The solution of the linearized equation~(\ref{qpblin}) is the polynomial background~(\ref{vbackground}). For odd $n=1,3,\dots$, it can be written as
\begin{eqnarray}
  v(x,t,n=2k+1) &=& -\frac{2^{2k+1} \lambda  (1-t)^k \Gamma \left(k+\frac{3}{2}\right)
   \, _1F_1\left(-k;\frac{1}{2};-\frac{x^2}{4
   (1-t)}\right)}{\sqrt{\pi }}\,,\quad k=0,1,\dots\,,
\end{eqnarray}
where $\, _1F_1\left(\dots\right)$ is the Kummer confluent hypergeometric function. 
For even $n=2,4,\dots$,  $v(x,t)$ is the following:
\begin{eqnarray}
    v(x,t,n=2k) &=& -\frac{2^{2k} k \lambda  x (1- t)^{k-1} \Gamma \left(k+\frac{1}{2}\right)
   \, _1F_1\left(1-k;\frac{3}{2};-\frac{x^2}{4(1-t)}\right)}{\sqrt{\pi}}\,,\quad k=1,2,\dots\,.
\end{eqnarray}
Using these results and Eq.~(\ref{rhobar}), we can calculate the linear-theory action  
\begin{equation}\label{slingenformula}
s(\lambda,n) \simeq \int_0^1 dt \int_{-\infty}^{\infty} dx \,\bar{u}^2(x,t) v^2(x,t,n)\,. 
\end{equation}
The result can be written in the form
\begin{equation}\label{slingenresult}
s(\lambda,n) \simeq\frac{a_n}{\sqrt{2\pi}}\,\lambda^2\,,
\end{equation}
where
\begin{equation}\label{anformula}
a_n=\frac{n^2[(n-1)!]^2}{2}\sum_{k=0}^{\lfloor(n-1)/2\rfloor}\frac{1}{(k!)^2\,(n-1-2k)!}\int_0^1 t^{\,n-2k-\frac32}\left(1-\frac t2\right)^{2k}dt
\end{equation}
is a rational number. The rationality follows by substituting $t=y^2$ in each integral.
The resulting integrand,
$2y^{2(n-1-2k)}(1-y^2/2)^{2k}$, is a polynomial with rational
coefficients, whose integral over $[0,1]$ is rational. Table~\ref{tab:an_values} shows the values of $a_n$ for the first several values of $n$.
\begin{table}[tb]
\caption{The constant $a_n$ in Eq.~(\ref{slingenresult}) for $n=1,\dots,8$.}
\label{tab:an_values}
\renewcommand{\arraystretch}{1.8}
\setlength{\tabcolsep}{7pt}
\begin{tabular}{ccccccccc}
\hline\hline
$n$ & 1 & 2 & 3 & 4 & 5 & 6 & 7 & 8 \\
\hline
$a_n$ & $1$ & $\dfrac{4}{3}$ & $\dfrac{147}{5}$ & $\dfrac{3888}{35}$ &
$\dfrac{57245}{21}$ & $\dfrac{1368060}{77}$ & $\dfrac{73208625}{143}$ &
$\dfrac{689956288}{143}$ \\[1mm]
\hline\hline
\end{tabular}
\end{table}
Using Eqs.~(\ref{shortcut}) and (\ref{slingenresult}), we obtain
\begin{equation}\label{slingenm}
  s(m_n) \simeq \sqrt{\frac{\pi}{8}} \,\frac{(\Delta m_n)^2}{a_n}\,,
\end{equation}
where $\Delta m_n \equiv m_n - \bar{m}_n$, and  $\bar{m}_n$ are given by Eq.~(\ref{barmn}).
Combining this quadratic rate function with Eq.~(\ref{SRW}), we obtain the long-time variances
\begin{equation}\label{momentvariances}
\operatorname{Var}(m_n)\simeq a_n\sqrt{\frac{2}{\pi T}}\,,\qquad
\operatorname{Var}(M_n)\simeq a_n\sqrt{\frac{2}{\pi}}\,W^2T^{n-\frac12}\,.
\end{equation}

\section{$n=1$: Duality to current statistics and the large-$|m_1|$ tail}\label{sec:n1}

There is a remarkably simple exact mapping, within MFT, between the statistics of the center of energy ($n=1$) for the delta-function initial density and the statistics of integrated current for a (deterministic) flat initial density. 
For $\lambda\ne0$, define
\begin{equation}\label{dualityfields}
U(x,t)=-\frac{v(x,1-t)}{\lambda}\,,\qquad
V(x,t)=\lambda u(x,1-t)\,.
\end{equation}
Direct substitution shows that $U,V$ obey the same MFT equations~(\ref{qpa}) and (\ref{qpb}), with boundary conditions
\begin{equation}\label{dualityboundaryconditions}
U(x,0)=1\,,\qquad V(x,1)=\lambda\delta(x)\,.
\end{equation}
The transformation preserves the action because $U^2V^2=u^2v^2$ at the reversed time.

To fix the sign convention, define the rescaled integrated current to the right by
\begin{equation}\label{dualitycurrent}
J=\int_0^\infty[U(x,1)-1]\,dx\,.
\end{equation}
Let $\nu$ denote the multiplier conjugate to $J$. With $V=-\partial_x P$, where $P$ is the conjugate field of the current problem, the terminal condition is $P(x,1)=\nu\theta(x)$, with the Heaviside step function $\theta$, i.e., $V(x,1)=-\nu\delta(x)$. Hence the mapping corresponds to $\nu=-\lambda$. Denoting the moment and current actions by $s_{\mathrm{moment}}$ and $s_{\mathrm{current}}$, respectively, their parametric representations satisfy $s_{\mathrm{moment}}(\lambda)=s_{\mathrm{current}}(-\lambda)$. The Legendre relations yield
\begin{equation}\label{dualityshortcut}
\lambda\frac{dm_1}{d\lambda}=-\lambda\frac{dJ}{d\lambda}\,.
\end{equation}
Both observables vanish in the deterministic limit $\lambda\to0$, so $J=-m_1$. Reflection symmetry of the flat-initial-density current problem therefore implies equality of the rate functions themselves:
\begin{equation}\label{dualityrates}
s_{\mathrm{moment}}(m_1)=s_{\mathrm{current}}(-m_1)=s_{\mathrm{current}}(m_1)\,.
\end{equation}
The quadratic actions obtained by linear theory in the two problems provide an immediate check of this identity: for a deterministic flat initial density, the variance of the integrated current $Q$ through the origin is $\langle Q^2\rangle\simeq\sigma(1)\sqrt{T/(2\pi)}=\sqrt{2T/\pi}$~\cite{DG2009b,KrMe}, so that $s_{\mathrm{current}}(J)\simeq\sqrt{\pi/8}\,J^2$, in agreement with Eq.~(\ref{s1linm}).

For large $|\lambda|$, corresponding to large $|m_1|$, we can therefore use the extreme-current asymptotic, which corresponds to a single adiabatically evolving doublon~\cite{MS2013}. After correcting an $O(1)$ error in their prefactors, as detailed in Appendix~\ref{app:currentcorrection}, the asymptotic results of Ref.~\cite{MS2013} take the parametric form
\begin{equation}\label{largelambda}
m_1(\lambda)\simeq\frac{e^{|\lambda|}}{\lambda\sqrt{\operatorname{Ei}(|\lambda|)}}\,,\qquad
s(\lambda)\simeq\frac{e^{|\lambda|}}{\sqrt{\operatorname{Ei}(|\lambda|)}}-2\sqrt{\operatorname{Ei}(|\lambda|)}\,,\qquad |\lambda|\gg1\,,
\end{equation}
where $\operatorname{Ei}(\dots)$ is the exponential integral function. These expressions satisfy $ds/d\lambda=\lambda\,dm_1/d\lambda$ and respect the odd parity of $m_1(\lambda)$ and the even parity of $s(\lambda)$.

\section{$n=2$: Distribution tails and dynamical phase transition}\label{sec:m2}

\subsection{The large-$m_2$ tail: An adiabatically evolving doublon}\label{sec:m2large}

As we verify below, at $m_2\gg1$, the optimal path breaks reflection symmetry: most of the energy is transported far from the origin by a single adiabatically evolving rapidly traveling doublon, as in the extreme-current problem of Ref.~\cite{MS2013}, see also Ref.~\cite{BSM2022b}.  We consider the right-moving  doublon; its reflected counterpart has the same second moment and action. The competition with the reflection-symmetric branch and the ensuing phase transition at $m_2=O(1)$ will be discussed in Sec.~\ref{sec:m2transition}.

The adiabatic perturbation theory \cite{MS2013} exploits the strong inequality $m_2\gg 1$, which is equivalent to the strong inequity $c \gg 1$ of the doublon velocity $c$. This strong inequality guarantees a length-scale separation between the small \emph{interior} region of the doublon and the large-scale \emph{exterior} regions in front of and behind the doublon. In the exterior regions the diffusion terms in the MFT equations (\ref{qpa}) and (\ref{qpb}) are negligible.  They are important, however, in the interior region, that is for the description of the doublon itself.  When viewed from the exterior regions, the doublon can be effectively described as a single \emph{point-like singularity} $X(t)$, which travels in space. The
singularity carries a finite energy $M(t)$, a finite jump of the ``momentum'' $V(t) = p_1 -  p_2$ (here the subscripts 1 and 2
refer to ``in front of the singularity'' and ``behind the singularity'', respectively), and a finite value of the Hamiltonian  $\mathcal{H}(t)$ of the MFT equations (\ref{qpa}) and (\ref{qpb}). In general, the quantities $M(t)$, $V(t)$ and $\mathcal{H}(t)$ slowly depend on time. They are defined as follows:
\begin{equation}\label{PM060}
    \{M(t),\,V(t),\,{\mathcal H}(t)\} = \int_{X(t)-\varepsilon}^{X(t)+\varepsilon} \{u,\,-v,\,h\}\, dx,
\end{equation}
where $h=v\,\partial_x u +u^2 v^2$, and $\varepsilon>0$ can be fixed at an arbitrary small value as $X\to\infty$ and $c \to \infty$.  Direct calculations~\cite{MS2013} show that
\begin{equation}\label{relations}
M(t)\simeq \frac{1}{\tilde{v}_1(t)}, \quad V(t)\simeq \frac{1}{u_2(t)} ,  \quad \text{and}\quad \mathcal{H}(t)\simeq c(t),
\end{equation}
where $\tilde{v}_1(t)$ is the value of $v$ in front of the doublon, and $u_2(t)$ is the value of $u$ behind the doublon. Conservation of $H$ (see Sec. \ref{sec:integrals}) implies that $X(t)=ct$ in the leading order. 

The doublon leaves a dilute energy wake and a slowly varying conjugate-field background, $v\simeq-2\lambda x$, behind it. We use this outer structure, supported by the numerical solutions, to describe the slow evolution of the pulse amplitudes.  Let $cA(t)$ denote the positive area of the narrow $-v$ pulse. This is the area of the pulse alone, with its smooth background excluded. The density pulse has width of order $c^{-1}$ and, locally, the profile \cite{MS2013}
\begin{equation}\label{m2profile}
u(x,t)\simeq\frac{cM(t)}{4}\operatorname{sech}^2\!\left[\frac{c(x-ct)}{2}\right].
\end{equation}
Applying the local forms of energy conservation and of the regularized conservation law \(\int_{-\infty}^{\infty}(v-v_{\mathrm{bg}})\,dx=0\), both discussed in Sec.~\ref{sec:integrals}, to an interval moving with the doublon yields, to leading adiabatic order,
\begin{equation}\label{m2amplitudes}
\dot M=-\frac{1}{A},\qquad \dot A=\frac{1}{M}-Bt,\qquad B=2\lambda c.
\end{equation}
The explicit time dependence comes from the background sampled by the moving doublon. The initial pulse carries the total mass, $M(0)=1$. For the outer solution we adopt the terminal prescription
\begin{equation}\label{m2endpoint}
M(1)=\frac{1}{B}.
\end{equation} 
This prescription represents matching to a short terminal layer; 
its accuracy and the reduction to Eq.~(\ref{m2amplitudes}) are discussed in Appendix~\ref{app:m2doublon}.

The second moment receives contributions both from the surviving pulse and from the wake. The mass $-\dot M\,dt$ is deposited near $x=ct$, and its subsequent displacement is negligible on the scale $c$. Consequently,
\begin{equation}\label{m2wake}
m_2=c^2M(1)+c^2\int_0^1t^2(-\dot M)\,dt
=2c^2\int_0^1tM(t)\,dt.
\end{equation}
Thus it is the entire history of mass loss, rather than the final pulse mass alone, that determines $m_2$. The local pulse solution gives the action accumulation rate
\begin{equation}\label{m2actionrate}
\dot s=c\{\ln(c^2MA)-2\}.
\end{equation}
This expression applies outside the short terminal layer. The finite term $-2c$ includes the contributions of both pulse cores and incorporates the prefactor correction derived in Appendix~\ref{app:currentcorrection}.

For each $B$, we solve Eqs.~(\ref{m2amplitudes}) and (\ref{m2endpoint}) with $M(0)=1$ and define
\begin{equation}\label{m2FG}
F(B)=2\int_0^1tM\,dt,\qquad
G(B)=\int_0^1[\ln(MA)-2]\,dt.
\end{equation}
Then $m_2=c^2F$ and $s_{\mathrm{ad}}=c(2\ln c+G)$. The shortcut relation~(\ref{shortcut}), with $\lambda=B/(2c)$, fixes the velocity and yields the parametric prediction
\begin{equation}\label{m2parametric}
c(B)=\exp\!\left(\frac{BF-G-2}{2}\right),\qquad
m_2(B)=c^2F,\qquad
s_{\mathrm{ad}}(B)=c(BF-2),\qquad
\lambda(B)=\frac{B}{2c}.
\end{equation}
The derivation, including the small endpoint term omitted at this order, is given in Appendix~\ref{app:m2doublon}. These expressions retain the full parameter dependence of the reduced amplitude problem and can be evaluated without expanding in $1/B$.

An explicit tail follows by taking $B\to\infty$. For any fixed $t<1$, $M(t)\to1$: the pulse transports nearly all the energy for most of the evolution, depositing it close to its final position. Although its final mass is only of order $B^{-1}$, the wake therefore has $m_2\simeq c^2$. Expanding the parametric solution gives
\begin{equation}\label{m2tail}
s(m_2)=\sqrt{m_2}\left[\ln m_2+\ln\ln m_2+\ln2-3+o(1)\right],\qquad m_2\to\infty.
\end{equation}
The corresponding doublon velocity and Lagrange multiplier are
\begin{equation}\label{m2clambda}
c=\sqrt{m_2}\left[1+\frac{1}{\ln m_2}+o\!\left(\frac{1}{\ln m_2}\right)\right],\qquad
\lambda=\frac{\ln m_2+\ln\ln m_2+\ln2-1+o(1)}{2\sqrt{m_2}}.
\end{equation}
The corrections in Eqs.~(\ref{m2tail}) and (\ref{m2clambda}) decay only as $\ln\ln m_2/\ln m_2$. For comparison with numerical MFT solutions at accessible moments, we use the full parametric prediction~(\ref{m2parametric}), which retains the finite terms of the reduced theory. Contributions not captured by the reduced theory, such as the formation of the doublon near $t=0$, the terminal layer, and higher-order adiabatic corrections,
may shift the action by $O(1)$ at accessible $m_2$; the offset observed numerically is discussed in Sec.~\ref{sec:numn2}.

\subsection{The small-$m_2$ tail}\label{sec:m2small}

A very small second moment requires almost all the energy to be concentrated near the origin at the observation time. We determine the leading action cost by bounding it from below and above using the solution of Bettelheim and Meerson~\cite{BM2024interval} for KMP energy confinement in a narrow interval with the same delta-function initial condition as ours. This approach does not require finding the optimal path for the second-moment constraint itself.

Let $\kappa=\int_{-\ell}^{\ell}u(x,1)\,dx$ be the energy fraction inside the interval, and let $s_\ell(\kappa)$ denote its rate function. We denote the Lagrange multiplier conjugate to $\kappa$ by $\mu>0$, to distinguish it from the second-moment multiplier $\lambda$.  For confinement of almost all the energy, \(\kappa\to1\), the narrow-interval solution of Ref.~\cite{BM2024interval}  gives
\begin{equation}\label{smallm2interval}
1-\kappa\simeq\frac{4\sqrt{2}}{\pi}\frac{\sqrt{\ln\mu}}{\mu},\qquad
s_\ell(\kappa)\simeq\frac{8\sqrt{2}}{3\pi}(\ln\mu)^{3/2},\qquad \text{where} \qquad
\mu\gg1,\quad \ell\sqrt{\ln\mu}\ll1.
\end{equation}
The numerical prefactors $O(1)$ here correct those in Eq.~(50) of Ref.~\cite{BM2024interval}; a short derivation is given in Appendix~\ref{app:smallm2}.

For any nonnegative terminal density, the energy outside the interval obeys the inequality
\begin{equation}\label{ineq10}
1-\kappa\leq\frac{1}{\ell^2}\int_{|x|>\ell}x^2u(x,1)\,dx
\leq\frac{m_2}{\ell^2}.
\end{equation}
Thus every path with the prescribed second moment $m_2$ also realizes an interval fraction $\kappa\geq\kappa_0\equiv1-m_2/\ell^2$.
This path, however, is suboptimal as far as the constraint on $\kappa$ is concerned. This argument yields a lower bound on $s(m_2)$ for any sufficiently small $\ell$:
\begin{equation}\label{smallm2lower}
s(m_2)\geq\inf_{\kappa\geq1-m_2/\ell^2}s_\ell(\kappa)
=s_\ell(1-m_2/\ell^2).
\end{equation}
Choose $\ell=[\ln(1/m_2)]^{-1}$. Equation~(\ref{smallm2interval}) then implies that, in the leading order, $\ln\mu \simeq \ln(1/m_2)$ at $m_2 \to 0$. As a result, $\ell\sqrt{\ln\mu}\to0$, consistently with the narrow-interval approximation of Ref.~\cite{BM2024interval}. Consequently, Eq.~(\ref{smallm2interval}) gives the leading-order asymptotic of the lower bound:
\begin{equation}\label{smallm2liminf}
\liminf_{m_2\to0}\frac{s(m_2)}{[\ln(1/m_2)]^{3/2}}
\geq\frac{8\sqrt{2}}{3\pi}.
\end{equation}
For an upper bound, the finite-interval paths of Ref.~\cite{BM2024interval} serve as admissible trial histories at their own second moments. They do obey energy conservation and the prescribed initial density and final moment, but they are suboptimal as far as the constraint on $m_2$ is concerned. In the limit of $\ell\to0$ at fixed $\mu$, the terminal density of Ref.~\cite{BM2024interval} consists of a core $\kappa\delta(x)$ and an exterior profile. The core contributes no second moment. Expanding the transform of the exterior profile, which is exact in this limit, gives (see Appendix~\ref{app:smallm2})
\begin{equation}\label{smallm2trialmoment}
m_2^{\mathrm{BM}}(\mu)\simeq
\frac{32\sqrt{2}}{3\pi^3}\frac{(\ln\mu)^{3/2}}{\mu},\qquad
\ln\frac{1}{m_2^{\mathrm{BM}}(\mu)}\simeq\ln\mu.
\end{equation} 
Returning to finite $\ell$, we assume that the action and the exterior second moment converge to their $\ell\to0$ limits locally uniformly in $\mu$. Convergence of the terminal density alone would not suffice to ensure convergence of its second moment. For each sufficiently large $\mu$, we can then retain a sufficiently small finite interval so that the relative errors in action and moment vanish as $\mu\to\infty$, while $\ell^2=o(m_2^{\mathrm{BM}})$. The core moment is at most $\kappa\ell^2$ and is therefore negligible. Since $m_2^{\mathrm{BM}}(\mu)$ decreases continuously and monotonically to zero as $\mu$ increases, for each sufficiently small prescribed moment we can choose $\ell$ small enough and then vary $\mu$ within the finite-interval family to attain that moment. Their action provides an upper bound whose leading-order asymptotic is given by Eq.~(\ref{smallm2interval}). Under the stated assumptions, the two bounds coincide to leading order, yielding
\begin{equation}\label{smallm2tail}
s(m_2)\simeq\frac{8\sqrt{2}}{3\pi}\left(\ln\frac{1}{m_2}\right)^{3/2},\qquad m_2\to0,
\end{equation}
which describes an asymptotic equivalence, rather than a pointwise bound at finite $m_2$. The subleading terms of the lower and upper bounds need not coincide.

For comparison with numerics, we retain the parametric dependence on $\mu$ in the leading large-$\mu$ expressions for the upper bound.
Unlike Eq.~(\ref{smallm2tail}), this procedure does not replace $\ln\mu$ by $\ln(1/m_2)$ and thus retains the logarithmic corrections generated by the implicit moment--multiplier relation. The corresponding slope is calculated as $\lambda=(ds/d\mu)/(dm_2^{\mathrm{BM}}/d\mu)$. This parametric approximation gives a more accurate comparison with numerics at accessible moments, as shown in Sec.~\ref{sec:numn2}. However, since both the action and the moment have been replaced by their leading large-$\mu$ expressions, the resulting curve need not remain an upper bound at finite $m_2$.

\subsection{Dynamical phase transition}\label{sec:m2transition}

When the second moment is close to its expected value $m_2=\bar m_2=2$, the optimal path  
is mirror-symmetric, as follows from the linear theory of Sec.~\ref{sec:Gn2}.
For sufficiently large $m_2$, however, the doublon mechanism favors
breaking this symmetry. To see this, we compare the single traveling
doublon of Sec.~\ref{sec:m2large} with a symmetric pair of doublons propagating
in opposite directions.

The single-doublon path has the leading action
$s_{\mathrm{one}}(m_2)\simeq\sqrt{m_2}\ln m_2$, see Eq.~(\ref{m2tail}).
In the symmetric path, each doublon carries half the initial energy
and contributes $m_2/2$ to the total second moment.
The MFT equations and action are invariant under
$u\to au$, $v\to v/a$. Normalizing each of the half-energy doublons to unit energy doubles its second moment from \(m_2/2\) to \(m_2\), while leaving its action unchanged. Each doublon therefore has the same leading action as the single unit-energy doublon realizing second moment \(m_2\). Neglecting subleading contributions from their
initial separation and subsequent interaction, their actions add:
\begin{equation}\label{twodoublons}
s_{\mathrm{two}}(m_2)\simeq 2s_{\mathrm{one}}(m_2)
\simeq 2\sqrt{m_2}\ln m_2.
\end{equation}
The symmetric two-doublon path is thus twice as expensive, to leading order, as either of the two single-doublon histories, which are mirror images of each other. Thus, at fixed large $m_2$, a single doublon carrying the full
energy is less costly than two counter-propagating doublons
sharing the energy equally.

Within the doublon description of the large-\(m_2\) tail, this comparison, together with the symmetric optimal path near \(m_2=2\), implies a symmetry-breaking dynamical phase transition at an intermediate value \(m_2^*\).
The argument does not determine the critical
moment or establish whether the transition is continuous or
discontinuous. These questions, including whether the symmetric
path remains a local minimum of the action at fixed $m_2$,
are addressed numerically in Sec.~\ref{sec:numn2}.

\section{Numerical solutions}

\subsection{Numerical algorithm}\label{sec:numericalalgorithm}

\paragraph*{Overview.} We numerically solve the MFT equations~(\ref{qpa}) and (\ref{qpb}) subject to the initial density condition~(\ref{incond1}) and the terminal conjugate-field condition~(\ref{fincond1}), with the prescribed moment fixed by Eq.~(\ref{constraint1}). A widely used numerical method is the back-and-forth iteration algorithm of Chernykh and Stepanov~\cite{CS2001}, see, e.g., Refs.~\cite{EK2004,KrMe,GGS2015,MKV2016,GBPC2017,BSM2022a,BSM2022b,AMPV2022,GB2023,BM2024interval}. In this method, one fixes the Lagrange multiplier $\lambda$ and alternately integrates the density equation~(\ref{qpa}) forward in time from the initial condition~(\ref{incond1}) and the conjugate-field equation~(\ref{qpb}) backward in time from the terminal condition~(\ref{fincond1}), using the latest available history of the other field. However, in the present problem we found that for sufficiently large $m_1$, and for $m_2$ sufficiently far from the expected value (in particular, in the large-$m_2$ and small-$m_2$ tails), these iterations converge poorly or fail entirely.

Therefore, instead of using the iteration algorithm, here we minimized the MFT action~(\ref{action}) directly, treating the noise of fluctuating hydrodynamics as a control field~\cite{GVE2019,SGMG2022,STGS2023,SSM2023,BSG2024,SG2026}. The gradient of the objective with respect to the control is obtained by the adjoint method~\cite{Plessix}, and the minimization uses a quasi-Newton method (L-BFGS) with line search~\cite{LN,NW}. Where the rate function is not convex, we impose the final-moment constraint~(\ref{constraint1}) through an augmented Lagrangian~\cite{Hestenes,Powell} rather than through a fixed multiplier $\lambda$. The three ingredients---the control formulation, the treatment of the constraint, and the discretization---are described in turn below.

\paragraph*{Control formulation.} In the rescaled variables, the fluctuating hydrodynamics equation~(\ref{Langevin}) of the KMP model reads
\begin{equation}\label{controlforward}
\partial_t u = \partial_x\left(\partial_x u + \sqrt{2}\,u\,\eta\right),\qquad u(x,0)=\delta(x)\,,
\end{equation}
where we now regard the noise history $\eta(x,t)$ as a deterministic \emph{control} field. Every $\eta$ produces a density history $u=u[\eta]$ and hence a value of the observable, $\mathcal{F}_n[\eta]=\int x^n u(x,1)\,dx$, at the price of the Gaussian action $s[\eta]=\frac12\int_0^1dt\int dx\,\eta^2$. The rate function is the minimal price for a prescribed moment,
\begin{equation}\label{controlproblem}
s(m_n)=\min_{\eta}\,s[\eta]\quad\text{subject to}\quad \mathcal{F}_n[\eta]=m_n\,.
\end{equation}
This is the same variational problem as in Sec.~\ref{sec:MFT}, written in terms of the noise. Indeed, enforcing Eq.~(\ref{controlforward}) with a Lagrange multiplier field $p(x,t)$ and the constraint with a scalar multiplier $\lambda$, the stationarity conditions of
\begin{equation}
\frac12\int_0^1\!\!dt\!\int\! dx\,\eta^2-\lambda\left(\mathcal{F}_n[\eta]-m_n\right)+\int_0^1\!\!dt\!\int\! dx\,p\left[\partial_t u-\partial_x\left(\partial_x u+\sqrt2\,u\eta\right)\right]
\end{equation}
are (i) $\eta=\sqrt2\,u v$ with $v=-\partial_x p$ (variation with respect to $\eta$), (ii) Eq.~(\ref{qpb}) for $v$ (variation with respect to $u$); and (iii) $p(x,1)=\lambda x^n$, i.e., the terminal condition~(\ref{fincond1}) (variation with respect to $u(x,1)$). Inserting $\eta=\sqrt2\,uv$ into Eq.~(\ref{controlforward}) gives Eq.~(\ref{qpa}), and $s[\eta]$ becomes the action~(\ref{action}). The ``adjoint field'' of optimal control is thus the conjugate field of MFT, and $\lambda$ is the multiplier of the shortcut relation~(\ref{shortcut}).

Instead of solving the resulting boundary-value problem directly, we minimize over $\eta$. The functional derivative of $s[\eta]-\lambda \mathcal{F}_n[\eta]$ is $\eta-\sqrt2\,u v$, where $u$ is obtained by integrating Eq.~(\ref{controlforward}) forward in time with the current $\eta$, and $v$ by integrating the (linear) conjugate-field equation backward in time from Eq.~(\ref{fincond1}), with the present $u$ and $\eta$. One gradient evaluation therefore costs one forward and one backward solve, independently of the number of degrees of freedom of $\eta$. The Chernykh--Stepanov iteration can be viewed as the fixed-point update $\eta\to\sqrt2\,uv$, i.e., a unit step along this gradient without any control of the step size. We instead pass the gradient to the quasi-Newton method L-BFGS~\cite{LN,NW}, which builds an approximation of the curvature of the objective from previous gradients and uses a line search that guarantees a decrease of the objective in every iteration. This is what makes the minimization robust in the strongly nonlinear doublon regime.

\paragraph*{Treatment of the constraint.} Where the rate function is convex, as for $n=1$, we minimize $s[\eta]-\lambda \mathcal{F}_n[\eta]$ at fixed $\lambda$. Each minimizer yields one point $\bigl(m_n=\mathcal{F}_n[\eta^*],\,s=s[\eta^*]\bigr)$, and $\lambda$ is changed in small steps, using the previous minimizer as initial guess. For $n=2$, the symmetry-broken branch is not convex: $\lambda=ds/dm_2$ decreases with $m_2$ (see Sec.~\ref{sec:numn2}). Global minimization at fixed $\lambda$ cannot select this branch, since the Legendre--Fenchel transform of the resulting tilted action retains only the convex envelope of $s(m_2)$~\cite{nonconvexnumerics}. We therefore impose the constraint through an augmented Lagrangian~\cite{Hestenes,Powell} and minimize
\begin{equation}\label{auglag}
s[\eta]-\lambda\left(\mathcal{F}_2[\eta]-m_2\right)+\frac{\gamma}{2}\left(\mathcal{F}_2[\eta]-m_2\right)^2
\end{equation}
for an increasing sequence of penalty parameters $\gamma$. After each minimization, the multiplier is updated as $\lambda\to\lambda-\gamma\,(\mathcal{F}_2[\eta^*]-m_2)$, and the next minimization starts from  the minimizer $\eta^*$ of the previous subproblem. The penalty adds a curvature $\gamma$ along the direction of the observable, which convexifies the problem locally as soon as $\gamma$ exceeds the magnitude of the negative curvature of $s(m_2)$, while the multiplier update drives $\lambda$ to $ds/dm_2$ at the prescribed~$m_2$.

\paragraph*{Discretization.} Following Refs.~\cite{STGS2023,SSM2023,BSG2024,SG2026}, we first discretize Eq.~(\ref{controlforward}) and then optimize. The objective is evaluated by a finite-volume solver (described below), and its gradient is computed by reverse-mode automatic differentiation of that solver in JAX~\cite{JAX}. This is equivalent to integrating the discrete backward (adjoint) equation, but needs no separate derivation or discretization of Eq.~(\ref{qpb}). The gradient is that of the discrete objective, up to rounding errors, as required by the line search. The conjugate field shown in the figures is obtained in the same way: $p$ is the derivative of the terminal term $\lambda \mathcal{F}_n$ with respect to the density at each time step, and $v=-\partial_x p$. An example implementation of this approach for a stochastic predator--prey model is provided in Ref.~\cite{SG2026}.

We pick a solver for Eq.~\eqref{controlforward} that is capable of adequately resolving doublons, i.e., narrow, fast pulses of $u$ and $v$, while remaining simple enough to be automatically differentiable. Writing Eq.~(\ref{controlforward}) as a conservation law, $\partial_t u+\partial_x j=0$, with flux $j=b u-\partial_x u$ and noise-induced drift $b=-\sqrt2\,\eta$, we use $N$ cells of width $\Delta x$ and zero-flux boundaries, so that the total energy is conserved exactly. The noise lives on the cell faces $x_{i+1/2}$. In the doublon, the optimal drift $b=-2uv$ is large, and the cell P\'eclet number $P_{i+1/2}=b_{i+1/2}\Delta x$ is not small (recall that the diffusion constant is $1$ in our choice of units). Central differences then produce spurious oscillations, whereas first-order upwinding smears out the pulses. We hence use the Scharfetter--Gummel or exponential-fitting flux~\cite{SG1969,Patankar},
\begin{equation}\label{SGflux}
j_{i+1/2}=\frac{1}{\Delta x}\left[\mathcal{B}(-P_{i+1/2})\,u_i-\mathcal{B}(P_{i+1/2})\,u_{i+1}\right],\qquad \mathcal{B}(z)=\frac{z}{e^z-1}\,,
\end{equation}
which follows from solving $b u-\partial_x u=j$ exactly between neighboring cell centers for constant $b$ and $j$. It reduces to central differences for $|P|\ll1$ and to upwinding for $|P|\gg1$. In time we use the Crank--Nicolson scheme with $n_t$ equidistant steps, holding $\eta$ constant during each step. The scheme is second-order accurate and unconditionally stable, and each step requires a single solve of a tridiagonal linear system, which is strictly diagonally dominant by columns, at a cost proportional to $N$. The initial condition~(\ref{incond1}) is replaced by a normalized Gaussian of width $\sigma_0=4\Delta x$. (At the linear level, this is equivalent to a time shift by $\sigma_0^2/2$, an $O(\Delta x)$ effect on the action that dominates the discretization error in the Gaussian regime below.) Finally, the terminal condition $v(x,1)=-n\lambda x^{n-1}$ is incompatible with our zero-flux numerical boundaries. We therefore replace $x^n$ in the observable by a smooth window function $w(x)$ which equals $x^n$ except within a distance $\delta=1$ from the boundaries, where it is continued by a cubic polynomial with $w'=0$ at the boundaries and continuous second derivative. As $u$ at the numerical boundaries is negligibly small, this does not affect the results.

The results below use $N=3000$ cells and $n_t=5000$ time steps. For $n=2$ the domain is $[-L/2,L/2]$ with $L=50$; for $m_2<1$ we use $L=15$, which resolves the energy concentrating near the origin. For $n=1$ the doublon moves to the right (for $\lambda>0$, without loss of generality), and we use the domain $[-L/4,3L/4]$ with $L=50$. Additional runs with $N=5000$ and $n_t=15000$ test the convergence. Each L-BFGS run is stopped when the relative change of the objective or the largest gradient component falls below $10^{-10}$, or after 5000 iterations. As an independent check, we monitored the integrals of motion $I_0$, $I_1$, and $I_2$ of Eqs.~(\ref{uv})--(\ref{I2}) along the computed histories.

\begin{figure}
\includegraphics[width=\textwidth]{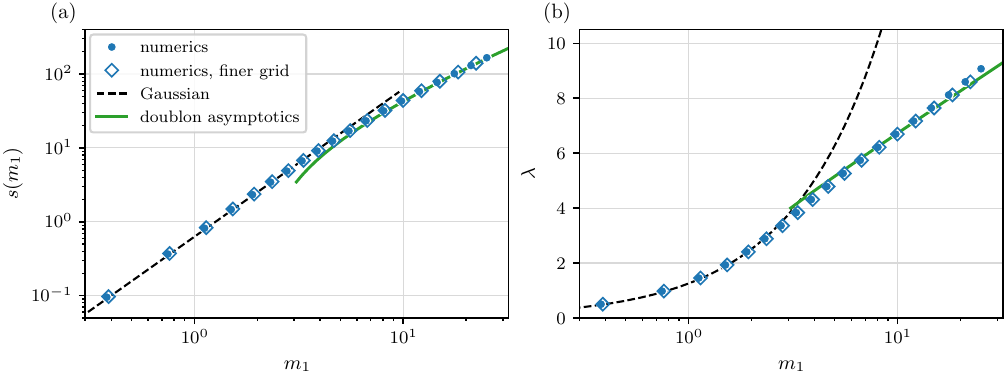}
\caption{First moment, $n=1$. (a) Rate function $s(m_1)$ and (b) Lagrange multiplier $\lambda=ds/dm_1$. Filled circles: numerical minimization with $N=3000$, $n_t=5000$; open diamonds: $N=5000$, $n_t=15000$. Dashed: linear theory, Eq.~(\ref{s1linm}), and $\lambda=\sqrt{\pi/2}\,m_1$. Solid: large-$\lambda$ doublon asymptotic, Eq.~(\ref{largelambda}).}
\label{fig:n1rate}
\end{figure}

\begin{figure}
\includegraphics[width=\textwidth]{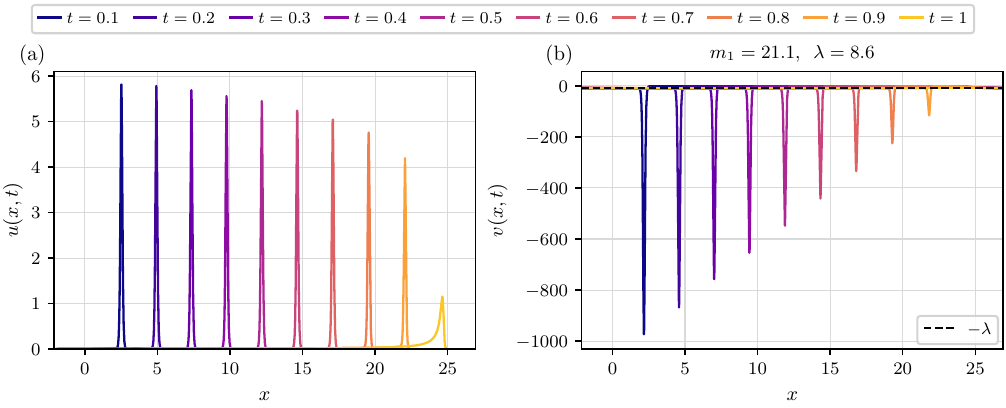}
\caption{Optimal path for $n=1$ at $\lambda=8.6$ in the doublon regime. (a) Density $u(x,t)$ and (b) conjugate field $v(x,t)$ at times $t=0.1,0.2,\dots,1$ (colors as in the legend). Dashed line in (b): the background $v=-\lambda$, Eq.~(\ref{vbackground}).}
\label{fig:n1doublon}
\end{figure}

\begin{figure}
\includegraphics[width=\textwidth]{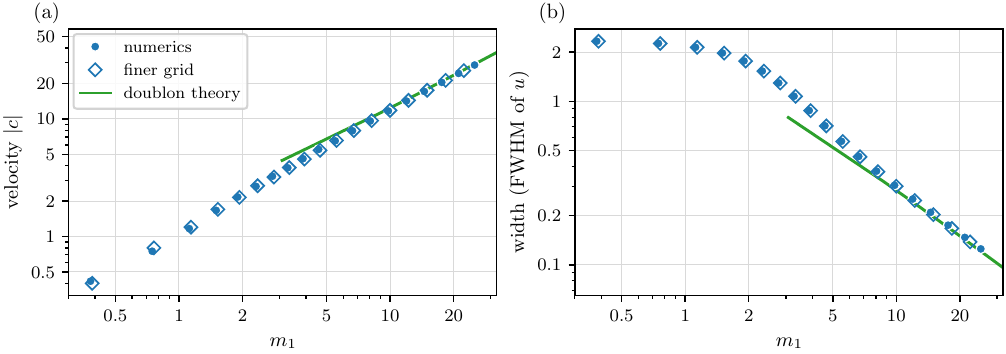}
\caption{Doublon (a) velocity and (b) width for $n=1$. The velocity is measured from the displacement of the maximum of $u(x,t)$ between $t=0.4$ and $t=0.6$; the width is the full width at half maximum of $u(x,t)$, averaged over $0.4\leq t\leq0.6$. Filled circles and open diamonds as in Fig.~\ref{fig:n1rate}. Solid lines: $c=[\operatorname{Ei}(|\lambda|)]^{1/2}$ along the parametric curve~(\ref{largelambda}), and the corresponding width $4\operatorname{arccosh}(\sqrt2)/c$ of the profile~(\ref{m2profile}).}
\label{fig:n1velocity}
\end{figure}

\begin{figure}
\includegraphics[width=\textwidth]{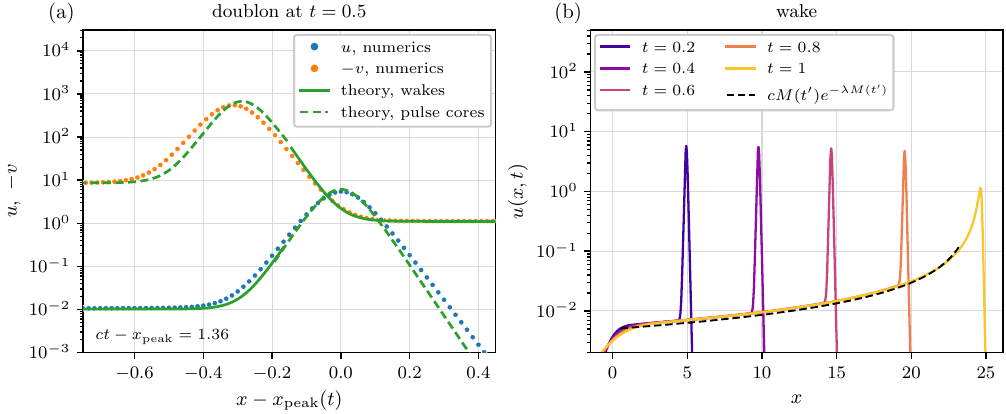}
\caption{Structure of the $n=1$ doublon at $\lambda=8.6$ (cf.\ Fig.~\ref{fig:n1doublon}). (a) Density $u$ (blue) and $-v$ (orange) at $t=1/2$, plotted against the distance from the numerical maximum $x_{\mathrm{peak}}(t)$ of $u$. Green curves: the adiabatic pulse solutions~(\ref{n1leading}) and (\ref{n1trailing}). Their parameters $c=[\operatorname{Ei}(\lambda)]^{1/2}$, $M=0.905$, $cA=98.0$ and $\xi_0=0.287$ come from Eqs.~(\ref{n1amplitudes}) and (\ref{n1separation}), with no fitting. Dashed: the $\operatorname{sech}^2$ pulses; solid: the monotonic parts of the solutions, which go over into the outer states. The theoretical profiles have been shifted to the left by $ct-x_{\mathrm{peak}}=1.36$, so that the maxima of $u$ coincide (see text). (b) Density $u(x,t)$ at $t=0.2,0.4,\dots,1$. Dashed: the theoretical wake~(\ref{n1wake}), with its horizontal coordinate adjusted using the numerical trajectory $x_{\mathrm{peak}}(t')\simeq24.4\,t'+0.03$, obtained from a linear fit over $0.2\leq t'\leq0.8$, and its amplitude $1/[cA(t')]$ unchanged (see text).}
\label{fig:n1profiles}
\end{figure}

\subsection{Theory vs. numerics for $n=1$}\label{sec:numn1}

Figure~\ref{fig:n1rate} shows the rate function and the Lagrange multiplier for the first moment. For $m_1\lesssim3$ the numerical results follow the linear theory, Eqs.~(\ref{slambdan=1}) and (\ref{s1linm}), within a few percent. For larger $m_1$ the numerical results cross over to the doublon asymptotics~(\ref{largelambda}). Via the duality of Sec.~\ref{sec:n1}, these are the extreme-current asymptotics of Ref.~\cite{MS2013} with the corrected prefactors of Appendix~\ref{app:currentcorrection}. Our results thus provide their first numerical confirmation. At the largest multipliers, the numerical $m_1(\lambda)$ and $s(\lambda)$ still depend on the resolution and approach Eq.~(\ref{largelambda}) under refinement (e.g.\ at $\lambda\simeq8.6$, the deviation of $m_1$ from Eq.~(\ref{largelambda}) decreases from $9\%$ to $4\%$). This reflects the narrowing of the doublon, whose width is proportional to $1/c$ and hence challenging to resolve, and, in part, the finite initial width $\sigma_0$, which is refined together with $\Delta x$.

The optimal path in Fig.~\ref{fig:n1doublon} shows the doublon that moves with nearly constant velocity and gradually loses energy to a dilute wake. It is accompanied by a sharp pulse of $v$ on the background $v=-\lambda$. Figure~\ref{fig:n1velocity} quantifies these observations. The measured velocity approaches the corrected prediction $c\simeq[\operatorname{Ei}(|\lambda|)]^{1/2}$, Eq.~(\ref{currentcorrected}), mapped by the duality.
The width agrees with the full width at half maximum (FWHM), $4\operatorname{arccosh}(\sqrt2)/c\simeq3.53/c$, of the $\operatorname{sech}^2$ profile~(\ref{m2profile}); for $\lambda\gtrsim7$, the product of measured width and measured velocity is $3.55$--$3.59$ on both grids.

Figure~\ref{fig:n1profiles} shows the structure of the doublon and its wake in more detail. Under the duality~(\ref{dualityfields}), combined with the reflection $x\to-x$ mapping $J<0$ onto the $j>0$ convention of Ref.~\cite{MS2013}, the density pulse of $u$ corresponds to the $v$ pulse of the current problem and the pulse of $-v$ to its $q$ pulse. The coupled-pulse solution of Ref.~\cite{MS2013}, their Eqs.~(65)--(70), then takes the following form. Let $\xi=x-ct$ be the coordinate comoving with the doublon, with the maximum of $u$ at $\xi=0$. Let $M(t)$ be the energy carried by the density pulse, and $cA(t)$ the area of the $-v$ pulse above its background $-v_{\mathrm{bg}}=\lambda$, as in Sec.~\ref{sec:m2large}. The leading density pulse and the region ahead of it are described by \cite{MS2013}
\begin{equation}\label{n1leading}
u\simeq\frac{cM}{4}\operatorname{sech}^2\frac{c\xi}{2}\,,\qquad
-v\simeq\frac{1}{M}\left(1+e^{-c\xi}\right),\qquad c(\xi+\xi_0)\gg1\,,
\end{equation}
which is Eq.~(\ref{m2profile}) for the density, accompanied by the exponential rise of $-v$ towards the trailing pulse. The trailing pulse of $-v$ and the region behind it are described by (with the background $\lambda$ restored)
\begin{equation}\label{n1trailing}
u\simeq\frac{1}{cA}\left[1+e^{c(\xi+\xi_0)}\right],\qquad
-v\simeq\lambda+\frac{c^2A}{4}\operatorname{sech}^2\frac{c(\xi+\xi_0)}{2}\,,\qquad -c\xi\gg1\,.
\end{equation}
Equations~(\ref{n1leading}) and (\ref{n1trailing}) go over into the local outer states $(u,-v)=(0,1/M)$ and $(1/cA,\lambda)$ immediately ahead of and behind the doublon, respectively. Both solutions hold in the overlap region between the two pulses. Matching them there determines the separation $\xi_0$ of the pulses [Eq.~(72) of Ref.~\cite{MS2013}]:
\begin{equation}\label{n1separation}
c\,\xi_0=\ln\left(c^2AM\right).
\end{equation}
The slow evolution of the amplitudes follows from Eq.~(\ref{mnamplitudes}) below with $n=1$, $B=\lambda$, or equivalently from Eqs.~(\ref{currentamplitudes}) with $k_1=c$ mapped by the duality. Using $MAe^{-\lambda M}=c^{-2}$ and $\dot M=-1/A$, we obtain
\begin{equation}\label{n1amplitudes}
\operatorname{Ei}[\lambda M(t)]=\operatorname{Ei}(\lambda)-c^2t\simeq c^2(1-t)\,,\qquad
cA(t)=\frac{e^{\lambda M(t)}}{cM(t)}\,,
\end{equation}
where the last step uses $c^2\simeq\operatorname{Ei}(\lambda)$. This determines all theoretical parameters for a given $\lambda$.

The predicted velocity $c\gg 1$ of the doublon carries an $O(1)$ subleading error. This error translates into an $O(1)$ error in the predicted position $ct$ of the doublon at time $t$. For example, for  $\lambda = 8.6$ we observed that the numerical doublon moves with velocity $24.4$ instead of the theoretically predicted $c=27.1$ (see Fig.~\ref{fig:n1velocity} and the resolution effects discussed above), so that at $t=1/2$ it lags behind $ct$ by $1.36$. For a fair comparison of the shapes, we therefore shift the theoretical profiles to the left by $ct-x_{\mathrm{peak}}$ so that the maxima of $u$ coincide. Figure~\ref{fig:n1profiles}(a) shows that, without fitting the pulse amplitudes, widths, or separation, Eqs.~(\ref{n1leading})--(\ref{n1amplitudes}) then capture the structure of the doublon fairly well over six decades.

Figure~\ref{fig:n1profiles}(b) addresses the wake of $u(x,t)$. As the density pulse sweeps through the point $x=ct'$, it loses energy at the rate $-\dot M=1/A$, Eq.~(\ref{mnamplitudes}) below, and spreads it over the distance $c\,dt'$. It thus leaves behind the density $1/[cA(t')]$, which is exactly the outer value of $u$ behind the doublon in Eq.~(\ref{n1trailing}). In the exterior region, the diffusion term in Eq.~(\ref{qpa}) acts on length scales $O(1)$, much smaller than the scale $c$ of the wake, and the nonlinear term is small because $u\ll1$. Therefore, the deposited energy essentially stays where it was left:
\begin{equation}\label{n1wake}
u(x,t)\simeq\frac{1}{cA(x/c)}=c\,M(x/c)\,e^{-\lambda M(x/c)}\,,\qquad 0<x<ct\,.
\end{equation}
As with the pulse formulas, this expression applies away from the initial and terminal layers, with $\lambda M(t')\gg1$ at the deposition time $t'=x/c$. Consistently, $\int_0^{ct}u\,dx=\int_0^t dt'/A(t')=1-M(t)$, as required by energy conservation. The wake density increases with $x$ because the doublon sheds energy faster as $M$ decreases. The numerical profiles $u(x,t)$ at different times indeed coincide behind the doublon, consistent with an approximately frozen wake. For the comparison in Fig.~\ref{fig:n1profiles}(b), we adjust only the horizontal coordinate of the theoretical wake using the measured trajectory, retaining the amplitude $1/[cA(t')]$. Specifically, we use $t'=(x-x_0)/c_{\mathrm{num}}$ instead of $x/c$, where $x_{\mathrm{peak}}(t)\simeq c_{\mathrm{num}}t+x_0$ with $c_{\mathrm{num}}=24.4$ and $x_0=0.03$ is a linear fit over $0.2\leq t\leq0.8$. This horizontal adjustment does not preserve the integrated wake energy; the energy-conservation identity above applies to the original Eq.~(\ref{n1wake}).

\begin{figure}
\includegraphics[width=\textwidth]{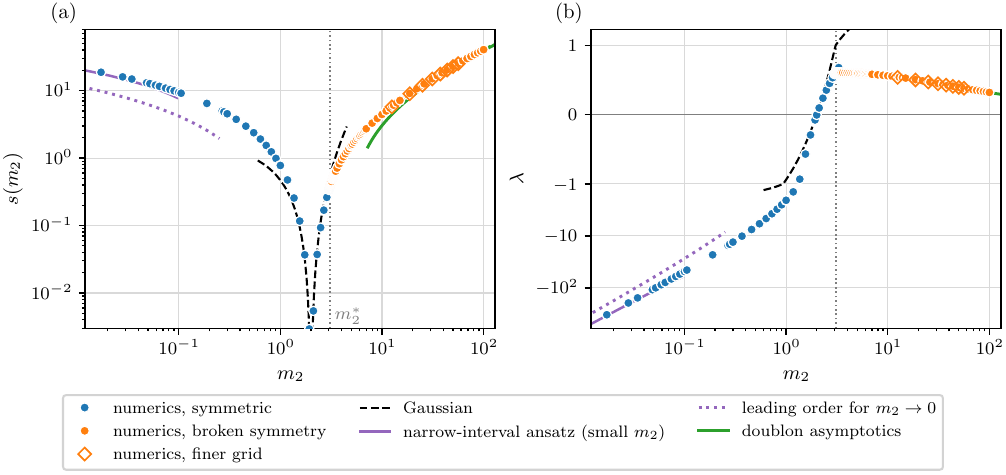}
\caption{Second moment, $n=2$. (a) Rate function $s(m_2)$ and (b) Lagrange multiplier $\lambda=ds/dm_2$ (symmetric-logarithmic scale). Circles: numerical minimization on the symmetric (blue) and symmetry-broken (orange) branches, with $L=15$ for $m_2<1$ and $L=50$ otherwise; open diamonds: finer grid. Dashed black curves near $\bar m_2=2$: linear theory, Eq.~(\ref{s2linm}). Solid purple: the leading large-$\mu$ action and moment of the upper-bound construction, Eqs.~(\ref{smallm2interval}) and (\ref{smallm2trialmoment}), evaluated parametrically as described in Sec.~\ref{sec:m2small}; this approximation need not remain an upper bound at finite $m_2$. Dotted purple: Eq.~(\ref{smallm2tail}), the common leading-order asymptotic of the lower and upper bounds, and its derivative in (b). Solid green lines at large $m_2$: the adiabatic doublon theory, Eq.~(\ref{m2parametric}). Vertical gray dotted line: the critical point $m_2^*\simeq3.07$.}
\label{fig:n2rate}
\end{figure}

\subsection{Theory vs. numerics for $n=2$}\label{sec:numn2}

Figure~\ref{fig:n2rate} shows on logarithmic axes the rate function and the multiplier for the second moment over almost four decades of $m_2$. Near the expected value $\bar m_2=2$ the numerical results follow the linear theory~(\ref{s2linm}). At $m_2^*\simeq3.07$ the optimal path breaks the reflection symmetry, and beyond it, $\lambda$ decreases with $m_2$. The rate function is hence concave on the symmetry-broken branch, which only the augmented Lagrangian minimization~(\ref{auglag}) can access thanks to its quadratic penalty term.

For large $m_2$, the multiplier agrees with the adiabatic doublon theory of Sec.~\ref{sec:m2large} to within about $1\%$ for $25\leq m_2\leq100$. The action exceeds $s_{\mathrm{ad}}$ of Eq.~(\ref{m2parametric}) by an offset of $1.29$--$1.38$ across $12.5\leq m_2\leq100$. A constant offset leaves $\lambda=ds/dm_2$ unchanged, consistent with the close agreement of the multipliers. The finer grid, used for $12.5\leq m_2\leq56.25$, lowers the action by about $1\%$. Since $s$ itself is large, this reduces the offset noticeably, to $1.31$--$1.09$, decreasing with $m_2$. Part of the offset is therefore a discretization effect.

For small $m_2$, the parametric approximation to the upper bound described in Sec.~\ref{sec:m2small} (solid purple curves in Fig.~\ref{fig:n2rate}) agrees well with the numerics. It uses the leading large-$\mu$ action and moment, Eqs.~(\ref{smallm2interval}) and (\ref{smallm2trialmoment}), without the further replacement $\ln\mu\simeq\ln(1/m_2)$. For example, at $m_2=0.0172$ it gives $s=17.8$ and $\lambda=-342$, compared with the numerical values $s=18.7$ and $\lambda=-330$, respectively. In contrast, Eq.~(\ref{smallm2tail}) (dotted purple curve) gives $s=9.83$, about a factor of two below the numerical action. Thus the logarithmic corrections retained by the parametric evaluation are substantial at accessible moments.

The solid purple curve approximates the upper-bound construction but is not itself a strict upper bound at finite $m_2$. Evaluating the unexpanded $\ell\to0$ trial-history expressions instead gives $s=24.5$ and a slope $-351$ at the same moment. Under the convergence assumptions of Sec.~\ref{sec:m2small}, this action supplies an upper bound, but its agreement with the numerical minimum is poorer: the interval-confinement history is suboptimal for the moment constraint. The closer agreement of the leading parametric approximation therefore reflects a partial compensation between its truncation error and the excess cost of the trial history.

\begin{figure}
\includegraphics[width=\textwidth]{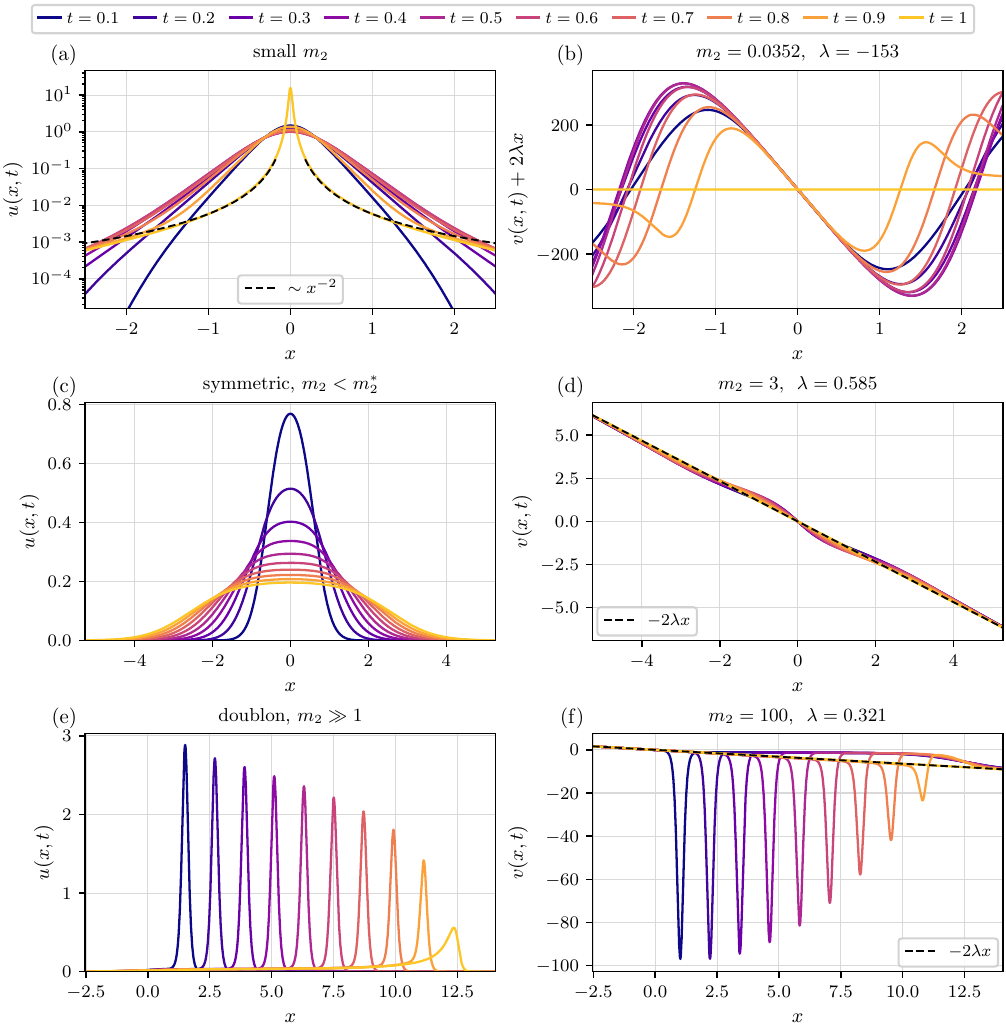}
\caption{Optimal histories for $n=2$ at times $t=0.1,0.2,\dots,1$. Left: density $u(x,t)$; right: conjugate field. (a,b) $m_2=0.035$: concentration of the energy near the origin. Dashed line in (a): $u(x,1)\sim x^{-2}$. In (b), $v$ is shown relative to its background, $v+2\lambda x$. (c,d) $m_2=3<m_2^*$: symmetric solution. (e,f) $m_2=100$: adiabatically evolving doublon (the right-moving one of the two mirror images). Dashed lines in (d,f): the background $v=-2\lambda x$, Eq.~(\ref{vbackground}).}
\label{fig:n2solutions}
\end{figure}

\begin{figure}
\includegraphics[width=\textwidth]{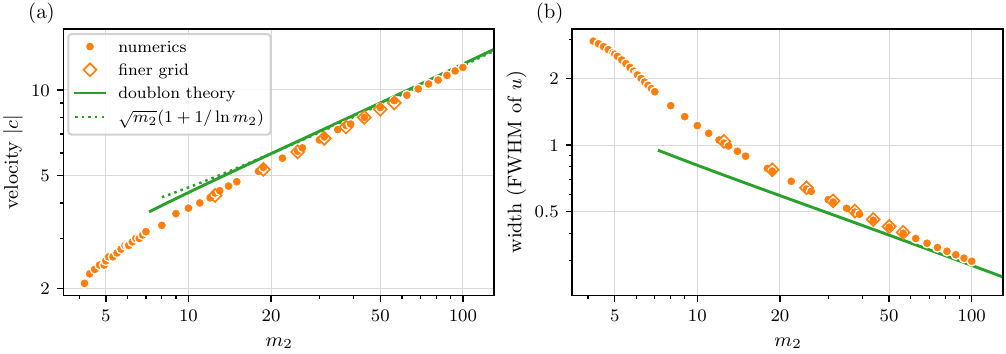}
\caption{Doublon (a) velocity and (b) width on the symmetry-broken branch for $n=2$, measured as in Fig.~\ref{fig:n1velocity}. Filled circles and open diamonds as in Fig.~\ref{fig:n2rate}. Solid lines: $c(B)$ from Eq.~(\ref{m2parametric}), and the corresponding width $4\operatorname{arccosh}(\sqrt2)/c$ of the profile~(\ref{m2profile}). Dotted in (a): $c$ from Eq.~(\ref{m2clambda}).}
\label{fig:n2velocity}
\end{figure}

\begin{figure}
\includegraphics[width=\textwidth]{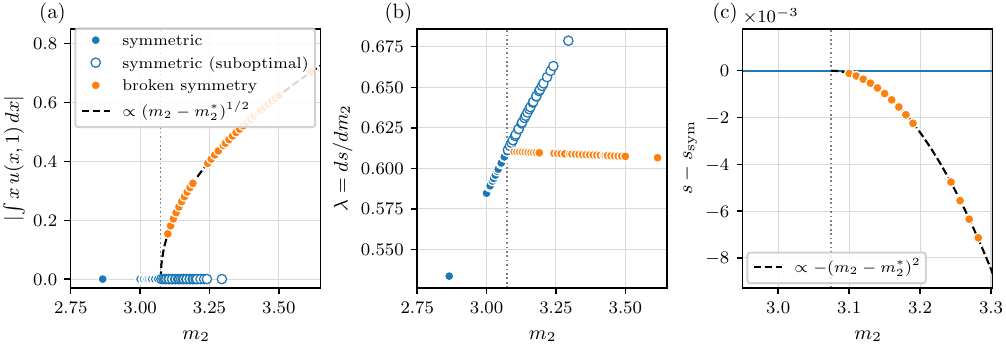}
\caption{Dynamical phase transition for $n=2$. (a) Order parameter $|m_1|=\left|\int x\,u(x,1)\,dx\right|$; dashed: fit $|m_1|^2=0.91\,(m_2-m_2^*)$, giving $m_2^*\simeq3.074$. (b) Lagrange multiplier. (c) Action difference between the broken and the symmetric branches; dashed: $-0.17\,(m_2-m_2^*)^2$. Filled (open) blue circles: symmetric solutions below (above) $m_2^*$; orange: symmetry-broken solutions.}
\label{fig:n2transition}
\end{figure}

Figure~\ref{fig:n2solutions} shows representative optimal histories. For small $m_2=0.035$ (top row) the energy first spreads and is then pulled back toward the origin, where it concentrates at $t=1$. Away from this core, the final density decays approximately as $u(x,1)\sim x^{-2}$ for $0.3\lesssim|x|\lesssim1.5$ and faster beyond.
For moderate $m_2=3$ (center row), slightly below the critical $m_2^*$, the optimal path is symmetric. Its energy spreads faster than in the deterministic solution~(\ref{rhobar}) and develops a flattened profile, while $v$ stays close to its background $-2\lambda x$. For large $m_2=100$ (bottom row) the optimal path is a doublon, as in Sec.~\ref{sec:m2large}. A narrow density pulse carries most of the energy away from the origin at nearly constant velocity, while depositing a dilute wake, and the sharp pulse of $v$ rides on the background $-2\lambda x$. Figure~\ref{fig:n2velocity} again compares the measured velocity and width with the adiabatic theory. At $m_2=100$ the velocity is $c=12.0$, compared with $12.3$ from Eq.~(\ref{m2parametric}) and $12.2$ from Eq.~(\ref{m2clambda}), and the width $0.299$ is close to the $\operatorname{sech}^2$ value $3.53/c$.

Finally, Fig.~\ref{fig:n2transition} resolves the transition region close to $m_2^*$. On the symmetry-broken branch, the order parameter $|m_1|=\left|\int x\,u(x,1)\,dx\right|$ vanishes as $|m_1|\sim(m_2-m_2^*)^{1/2}$, with $m_2^*\simeq3.074$ determined numerically. The multiplier $\lambda=ds/dm_2$ is continuous at $m_2^*$, $\lambda\simeq0.610$, but its slope $d^2s/dm_2^2$ jumps. Correspondingly, the action difference between the broken and the symmetric branches is $\Delta s\simeq-0.17\,(m_2-m_2^*)^2$. Since both branches share $s$ and $\lambda$ at $m_2^*$, this agrees with the estimate $\Delta s\simeq\tfrac12\,\bigl[(d^2s/dm_2^2)_{\mathrm{broken}}-(d^2s/dm_2^2)_{\mathrm{sym}}\bigr](m_2-m_2^*)^2\simeq-0.17\,(m_2-m_2^*)^2$ obtained from the jump of the second derivative. The transition is therefore of second order, with a mean-field (Landau) order-parameter exponent $1/2$. Beyond $m_2^*$ the symmetric solution, computed from mirror-symmetric initial guesses, continues to exist, but it is no longer a minimum of the action. A numerical analysis of the second variation of the action around it~\cite{STGS2023,SG2026} shows a single negative eigenvalue appearing at $m_2^*$, with an eigenfunction that breaks the reflection symmetry.

\section{Higher moments}\label{sec:higher}

The Gaussian regime, described by linear theory, was treated for all $n$ in Sec.~\ref{sec:Gaussian}. Here we show that the analysis of the tails for $n=1$ and $2$ extends to arbitrary $n$ with few changes.

\paragraph*{Large $|m_n|$.} By the invariance $u\to au$, $v\to v/a$ (Sec.~\ref{sec:m2transition}), a doublon carrying a fraction $a<1$ of the energy to the moment $m_n$ costs as much as one carrying the whole energy to the larger moment $m_n/a$. Within the doublon mechanism it is therefore cheapest to transport the whole energy with a single doublon. At $x\simeq ct$ with $c\gg1$, the background~(\ref{vbackground}) is dominated by its leading term, $-v_{\mathrm{bg}}\simeq n\lambda(ct)^{n-1}$, so that the reduced amplitude equations~(\ref{m2amplitudes}) and the terminal prescription~(\ref{m2endpoint}) become
\begin{equation}\label{mnamplitudes}
\dot M=-\frac1A,\qquad \dot A=\frac1M-Bt^{n-1},\qquad M(0)=1,\qquad M(1)=\frac1B,\qquad B=n\lambda c^{n-1}
\end{equation}
(we assume the right-moving doublon, $m_n>0$ and $\lambda>0$). The wake contributes as in Eq.~(\ref{m2wake}), now giving $m_n=c^nF$ with $F=n\int_0^1t^{n-1}M\,dt$, while the action rate~(\ref{m2actionrate}) and $G$ of Eq.~(\ref{m2FG}) are unchanged. The identity~(\ref{m2identity}) holds with $BF'/2$ replaced by $BF'/n$, and Eq.~(\ref{m2parametric}) generalizes to
\begin{equation}\label{mnparametric}
c=\exp\!\left(\frac{BF-G-2}{2}\right),\qquad
m_n=c^nF,\qquad
s_{\mathrm{ad}}=c\,(BF-2),\qquad
\lambda=\frac{B}{n\,c^{n-1}}.
\end{equation}
For $n=1$, $B=\lambda$ is constant, and Eq.~(\ref{mnamplitudes}) can be integrated in closed form: $MAe^{-BM}=\mathrm{const}$, $c^2=\operatorname{Ei}(\lambda)-\operatorname{Ei}(1)$, $m_1=(e^\lambda-e)/(\lambda c)$ and $s=\lambda m_1-2c$. This reproduces Eq.~(\ref{largelambda}) up to relative corrections $O(\lambda e^{-\lambda})$, which provides an independent check both of the reduced theory and of the corrected prefactors of Appendix~\ref{app:currentcorrection}. For arbitrary $n$, the large-$B$ analysis of Appendix~\ref{app:m2doublon} gives $A\simeq(B/n)(1-t^n)$, $F=1-n/B+O(B^{-2})$ and $G=\ln(B/n)-H_{1/n}-2+o(1)$, where $H_{1/n}=-\int_0^1\ln(1-t^n)\,dt=\psi(1+1/n)+\gamma_E$ is a harmonic number ($H_1=1$, $H_{1/2}=2-2\ln2$). Hence
\begin{equation}\label{mntail}
s(m_n)=|m_n|^{1/n}\left[\frac2n\ln|m_n|+\ln\ln|m_n|+\ln\frac{2}{n^2}-H_{1/n}-1+o(1)\right],\qquad |m_n|\to\infty,
\end{equation}
with $c\simeq|m_n|^{1/n}[1+n/(2\ln|m_n|)]$. For $n=2$ these reduce to Eqs.~(\ref{m2tail}) and (\ref{m2clambda}). In terms of the distance $c\simeq|m_n|^{1/n}$ traveled by the doublon, the leading action $2c\ln c$ is independent of $n$. For all $n\geq2$,
the displayed asymptotic is concave, so that the rate function is concave in the large-$|m_n|$ tail. The corrections to Eq.~(\ref{mntail}) decay slowly, and the explicit form becomes accurate only at very large $|m_n|$; at accessible moments one should use Eq.~(\ref{mnparametric}).

\paragraph*{Odd and even $n$.} For odd $n$, reflection $x\to-x$ maps $m_n\to-m_n$, so that $s(m_n)=s(-m_n)$, and both tails are realized by a single doublon moving in the direction of $\operatorname{sign}m_n$. There is no symmetry to break at fixed $m_n$, and, as for $n=1$, we expect the optimal path to evolve continuously from the Gaussian regime to the doublon. For even $n$, the comparison of Sec.~\ref{sec:m2transition} applies verbatim: a symmetric pair of half-energy doublons costs twice the action of a single doublon, which implies a symmetry-breaking transition at some $m_n^*$. We have verified this numerically for $n=4$, using the method of Sec.~\ref{sec:numericalalgorithm} on a coarse grid ($N=1000$, $n_t=1500$, $L=40$).
As for $n=2$, our numerical results indicate that the transition is of second order. On the symmetry-broken branch, the squared order parameter $|m_1|^2$ grows linearly with $m_4-m_4^*$, $\lambda$ is continuous at $m_4^*$, and $d^2s/dm_4^2$ jumps. Extrapolating the numerics to initial width $\sigma_0 = 0$ gives the estimate $m_4^*\simeq20.0$ for the critical $m_4^*$.
At larger $m_4$, the numerical $\lambda = \lambda(m_4)$ approaches the prediction of Eq.~(\ref{mnparametric}), with deviations of $5\%$, $2\%$, and $0.7\%$ at $m_4=70$, $150$, and $250$, respectively, while the action exceeds $s_{\mathrm{ad}}$ by a nearly constant offset of about $1.5$, similar to what was found for $n=2$.

\paragraph*{Small $m_n$ for even $n$.} We take $m_n\to0$ at fixed even $n$. The lower bound of Sec.~\ref{sec:m2small} carries over with $1-\kappa\leq m_n/\ell^n$, under the same assumptions on the interval-confinement solution. For the upper bound, expanding the transform of its exterior terminal density gives (see Appendix~\ref{app:smallm2})
\begin{equation}\label{smallmntrialmoment}
m_n^{\mathrm{BM}}(\mu)\simeq\frac{2\mathcal D^{n+1}}{(n+1)\mu}
\simeq\frac{2}{n+1}\left(\frac{2\sqrt2}{\pi}\right)^{n+1}
\frac{(\ln\mu)^{(n+1)/2}}{\mu},
\end{equation}
where $\mathcal D$ is defined in Appendix~\ref{app:smallm2}. This is the moment of the leading plateau of height $1/\mu$ on $|x|<\mathcal D$ and reproduces Eq.~(\ref{smallm2trialmoment}) for $n=2$. For comparison with numerics at accessible $m_n$, one can retain
the parametric dependence on $\mu$ in the action,
Eq.~(\ref{smallm2interval}), and the moment,
Eq.~(\ref{smallmntrialmoment}), rather than replacing
$\ln\mu$ by $\ln(1/m_n)$. As for $n=2$, this retains logarithmic
corrections but need not preserve the upper-bound property
at finite moments.

In returning to finite intervals, we now assume that the action and the exterior \emph{$n$th moment} converge to their $\ell\to0$ limits locally uniformly in $\mu$; convergence of the second moment alone would not suffice. Choosing $\ell^n=o(m_n^{\mathrm{BM}})$ makes the core contribution, bounded by $\kappa\ell^n$, negligible. The moment can then be matched within the finite-interval family as in Sec.~\ref{sec:m2small}. Since $\ln[1/m_n^{\mathrm{BM}}(\mu)]\simeq\ln\mu$, the upper and lower bounds give, under these assumptions, the leading-order asymptotics which, remarkably, is $n$-independent: 
\begin{equation}\label{smallmntail}
s(m_n)\simeq\frac{8\sqrt{2}}{3\pi}\left(\ln\frac{1}{m_n}\right)^{3/2},\qquad m_n\to0,\quad n\ \text{even}.
\end{equation}

\section{Summary and discussion}\label{sec:discussion}

We have studied the long-time statistics of spatial moments of the energy distribution in the freely evolving KMP model following a localized energy release. These observables provide a family of correlated sums whose statistics are generated by the nonequilibrium dynamics. MFT relates their probability distributions to the optimal paths of energy transport and redistribution. For every positive integer \(n\), we calculated the leading variance, \(\operatorname{Var}(M_n)\sim W^2T^{n-1/2}\), including its \(n\)-dependent coefficient. Consequently, the typical fluctuations of the rescaled moments \(m_n=M_n/(WT^{n/2})\) decrease as \(T^{-1/4}\), independently of \(n\).

The first moment admits an exact duality, within MFT, to the integrated current for a deterministic flat initial density. This duality identifies the rate functions of two fluctuation problems with very different initial conditions and observables. It also transfers the adiabatic doublon description of extreme currents to large displacements of the center of energy. Our numerical solutions support this description and, through the duality, provide the first numerical confirmation of the extreme-current asymptotics, with the corrected prefactors.

For the second moment, the two distribution tails correspond to different physical mechanisms. A very large second moment is realized by a single traveling doublon that transports energy far from the release point and deposits it in a wake. Its slow evolution yields the leading action \(s(m_2)\simeq\sqrt{m_2}\ln m_2\), together with a logarithmic correction and a parametric approximation useful for comparison with numerics. A very small second moment requires almost complete concentration of the energy near the origin at the observation time. Using the interval-confinement solution of Bettelheim and Meerson, we obtained lower and upper bounds that coincide to leading order and, under the assumptions stated in Sec.~\ref{sec:m2small}, give \(s(m_2)\simeq[8\sqrt{2}/(3\pi)][\ln(1/m_2)]^{3/2}\). This approach determines the leading action without requiring the complete optimal path for the second-moment constraint.

A central finding is the dynamical phase transition in the second-moment statistics. Although both the initial condition and the second-moment constraint are reflection-symmetric, a sufficiently large second moment is achieved more economically by transporting energy predominantly in one direction. The two resulting optimal paths are mirror images of each other and have opposite first moments. Comparing single- and two-doublon paths predicts symmetry breaking, while the numerical solutions locate the transition at \(m_2^*\simeq3.07\) and establish its second-order character. The order parameter vanishes as \(|m_1|\sim(m_2-m_2^*)^{1/2}\), and the second derivative of the rate function is discontinuous at the transition.

Direct minimization of the MFT action made it possible to investigate these strongly nonlinear regimes numerically. Treating the noise history as a control field avoids the convergence difficulties of the back-and-forth iteration method, while imposing the moment constraint through an augmented Lagrangian gives access to the nonconvex rate function. The comparisons also demonstrate the practical importance of retaining parametric asymptotic formulas: slowly varying logarithmic corrections can remain substantial throughout the numerically accessible range.

The analysis extends to higher moments (Sec.~\ref{sec:higher}). For every \(n\), the large-\(|m_n|\) tail is governed by a single adiabatically evolving doublon, with the leading action \(s\simeq(2/n)|m_n|^{1/n}\ln|m_n|\). For all even \(n\), this implies a reflection-symmetry-breaking transition, which we found to be of second order also for \(n=4\), and the leading small-\(m_n\) tail is the same as for \(n=2\). Determining the critical moments \(m_n^*\) for general even \(n\) remains open. The joint statistics of the first and second moments would distinguish fluctuations of the center of energy from fluctuations of the spreading about that center, providing a fuller description of the fluctuating energy packet.

For \(n=1\), the terminal condition \(v(x,1)=-\lambda\) is spatially uniform, and, by the duality of Sec.~\ref{sec:n1}, the problem is equivalent to the integrated-current problem for a flat deterministic initial density. It is natural to attempt to solve this problem exactly for all $m_1$ by using the ISM \cite{BSM2022a,MMS2022,BM2024interval}, adapted for fields that do not decay at infinity. To date, however, such attempts have proven unsuccessful. For \(n\geq2\), the polynomial growth of \(v\) at large \(|x|\) appears to place the problem beyond the standard ISM framework.

A further interesting question concerns possible singular concentration histories. Heydecker~\cite{Heydecker2026} has recently shown that, even in one dimension, smooth KMP histories with bounded action can approach a history with point-mass concentration at an isolated time. His construction retains a positive background on a periodic domain and does not establish a competing history for our localized initial condition and moment constraints. (On the lattice, the formal density singularity must be regularized
by concentration on one or a few sites, whose energies remain
bounded by the conserved total energy.) Whether such singular limits affect the global minimization of the action in our setting --- and in previously studied settings where the MFT was employed --- remains open.

\begin{acknowledgments}
We are grateful to Arkady Vilenkin for his participation in the initial stage of this project. The research of T.S.\ was supported in part by the International Centre for Theoretical Sciences (ICTS) for participating in the program - Two and a half decades of the Macroscopic Fluctuation Theory 2026 (code: ICTS/MFTHEORY{\allowbreak}2026/09).
The research of B.M.\ is supported by the Israel Science Foundation (Grant
No.~1579/25) and was also supported in part by NSF Grant No.~PHY-2309135 to the
Kavli Institute for Theoretical Physics (KITP). The authors acknowledge the use of generative AI to assist in some of the derivations, coding, and manuscript drafting. The AI output was reviewed and verified by the authors, who take full responsibility for the results presented in this work.
\end{acknowledgments}

\appendix
\section{Correcting an $O(1)$  prefactor error in the extreme-current asymptotics.}\label{app:currentcorrection}

The prefactor correction used in Sec.~\ref{sec:n1} originates from two omitted subleading contributions (one from each pulse core) to the doublon action in Appendix B.3 of Ref.~\cite{MS2013}. Here we recover these contributions and their consequences. Throughout this appendix, $q$, $v$, $M$, $V$, and $\Lambda$ follow the notation of that reference: $q$ is the energy density, $v$ is the conjugate-field gradient, $M$ and $V$ are the areas of the density and conjugate-field pulses, respectively, and $\Lambda$ is the rescaled current multiplier (not to be confused with $\Lambda$ of Sec.~\ref{sec:MFT}). In particular, the sign convention for $v$ differs from ours. All subsequent relations are understood within the adiabatic approximation.

Let $c>0$ be the doublon velocity and $\xi_0$ the separation between the trailing conjugate-field pulse and the leading density pulse. We place the trailing pulse at $\xi=0$ and choose a matching point $\xi_1$ such that $c\xi_1\gg1$ and $c(\xi_0-\xi_1)\gg1$. The trailing-pulse solution gives $q_<v_<=ce^{c\xi}/(1+e^{c\xi})$, so its contribution to the action rate is
\begin{equation}\label{currenttrailing}
\int_{-\infty}^{\xi_1}q_<^2v_<^2\,d\xi
=c\int_0^{e^{c\xi_1}}\frac{y}{(1+y)^2}\,dy
=c^2\xi_1-c+O\!\left(ce^{-c\xi_1}\right).
\end{equation}
Similarly, the leading pulse contributes
\begin{equation}\label{currentleading}
\int_{\xi_1}^{\infty}q_>^2v_>^2\,d\xi
=c^2(\xi_0-\xi_1)-c
+O\!\left(ce^{-c(\xi_0-\xi_1)}\right).
\end{equation}
The terms $-c$ were omitted in Eqs.~(B.6) and (B.7) of Ref.~\cite{MS2013}. Restoring them and using $c\xi_0=\ln(cMV)$ yields
\begin{equation}\label{currentactionrate}
\dot s=c^2\xi_0-2c=c\left[\ln(cMV)-2\right].
\end{equation}

The amplitude equations remain unchanged:
\begin{equation}\label{currentamplitudes}
\dot M=c\left(1-\frac{1}{V}\right),\qquad
\dot V=\frac{c}{M},\qquad
M=\frac{e^V}{k_1(c)V},
\end{equation}
where $k_1(c)$ is an integration constant. At the final time, $V(1)=\Lambda$ and $M(1)=j$, where $j>0$ is the rescaled current ($|J|$ in Sec.~\ref{sec:n1}). Neglecting the short initial stage, as in Ref.~\cite{MS2013}, integration gives
\begin{equation}\label{currentintegrated}
\operatorname{Ei}(\Lambda)=ck_1,\qquad
j=\frac{e^\Lambda}{\Lambda k_1},\qquad
s=\Lambda j+c\ln\frac{c}{k_1}-2c.
\end{equation}
The first two relations imply
\begin{equation}\label{currentvariation}
j\,d\Lambda=dc+\frac{c}{k_1}\,dk_1.
\end{equation}
Differentiating the action in Eq.~(\ref{currentintegrated}) therefore gives
\begin{equation}\label{currentdifferential}
ds-\Lambda\,dj=\ln\frac{c}{k_1}\,dc.
\end{equation}
The identity $ds/dj=\Lambda$ fixes $k_1=c$, instead of $k_1=e^2c$ obtained in Ref.~\cite{MS2013}. Consequently, the corrected parametric result is
\begin{equation}\label{currentcorrected}
c=\sqrt{\operatorname{Ei}(\Lambda)},\qquad
j=\frac{e^\Lambda}{\Lambda\sqrt{\operatorname{Ei}(\Lambda)}},\qquad
s=\Lambda j-2c.
\end{equation}
At fixed $\Lambda$, the corrected $c$, $j$, and $s$ are each larger by a factor of $e$ than the expressions in Ref.~\cite{MS2013}. The duality of Sec.~\ref{sec:n1}, with $j=|m_1|$ and $\Lambda=|\lambda|$, gives Eq.~(\ref{largelambda}).
Although the omitted term $-2c$ is subleading in the action rate, it is of the same order as the term $c\ln(c/k_1)$ that determines $k_1$. It must therefore be retained to obtain the correct prefactors.

\section{Amplitude reduction and asymptotics of the second-moment doublon}\label{app:m2doublon}

The outer states immediately ahead of and behind the right-moving doublon are, respectively,
\begin{equation}\label{m2outerfields}
(u_1,-v_1)=\left(0,\frac{1}{M}\right),\qquad
(u_2,-v_2)=\left(\frac{1}{cA},B t\right).
\end{equation}
Applying the moving-interval balance to Eqs.~(\ref{qpa}) and (\ref{qpb}), and retaining the nonlinear fluxes of the outer fields, gives
\begin{equation}\label{m2balancefull}
\dot M=-\frac{1}{A}+\frac{2Bt}{c^2A^2},\qquad
\dot A=\frac{1}{M}-Bt+\frac{2B^2t^2}{c^2A}.
\end{equation}
Away from the terminal layer, the terms proportional to $c^{-2}$ can be dropped, yielding Eq.~(\ref{m2amplitudes}). The diffusive fluxes of the outer fields, neglected in Eq.~(\ref{m2balancefull}), contribute at the same order; this equation serves only to estimate where the reduction fails, namely for $A=O(B/c^2)$. The sharp conjugate-field pulse is accompanied by a broad compensating component; its area $cA$ should not be identified with the integral of the entire background-subtracted field.

Near $t=1$, replacing $t$ by unity in the second reduced amplitude equation gives $d\ln A/dM=B-1/M$, or $MAe^{-BM}=\mathrm{const}$. Matching to the outer solution yields
\begin{equation}\label{m2matching}
A(M)\simeq\frac{2B}{M}e^{-B(1-M)}.
\end{equation}
Together with $c^2\simeq e^B/(2B)$, derived below, this implies $A=O(Bc^{-2})$ when $M=O(B^{-1})$. The terminal layer has duration $O(c^{-2})$. Equation~(\ref{m2endpoint}) selects a convenient matching point within this regime. Uncertainty in this matching does not affect the orders retained in Eq.~(\ref{m2tail}).

To derive Eq.~(\ref{m2parametric}), differentiate the reduced amplitude equations with respect to $B$. They imply
\begin{equation}\label{m2amplitudevariation}
\frac{d}{dt}(A\partial_B M)=\partial_B\ln(MA)-Bt\partial_B M.
\end{equation}
Using the endpoint conditions and Eq.~(\ref{m2FG}), we obtain
\begin{equation}\label{m2identity}
G'-\frac{B}{2}F'=-\frac{A(1)}{B^2}.
\end{equation}
On the other hand, $ds/dB=\lambda\,dm_2/dB$ gives
\begin{equation}\label{m2speedvariation}
\frac{d\ln c}{dB}\,(2\ln c+G+2-BF)=\frac{B}{2}F'-G'.
\end{equation}
Neglecting the endpoint term in Eq.~(\ref{m2identity}) consistently with the amplitude reduction gives $2\ln c=BF-G-2$ on the large-$B$ branch.

To obtain the explicit asymptotics, we need only the large-$B$ expansions
\begin{equation}\label{m2expansion}
\begin{aligned}
M(t)&=1-\frac{1}{B}\ln\frac{1+t}{1-t}+O(B^{-2}),\\
A(t)&=\frac{B}{2}(1-t^2)+O(1),
\end{aligned}
\end{equation}
valid at fixed $t<1$. Matching to the endpoint region and evaluating the integrals in Eq.~(\ref{m2FG}) gives, to the required orders,
\begin{equation}\label{m2FGexpansion}
F=1-\frac{2}{B}+O(B^{-2}),\qquad
G=\ln(2B)-4+O(B^{-1}).
\end{equation}
Substitution into Eq.~(\ref{m2parametric}) yields
\begin{equation}\label{m2logspeed}
2\ln c=B-\ln(2B)+O(B^{-1}),
\end{equation}
and hence $c^2\simeq e^B/(2B)$. Since $m_2=c^2F$, we have $B=\ln m_2+\ln\ln m_2+\ln 2+o(1)$. Using $s_{\mathrm{ad}}/\sqrt{m_2}=(BF-2)/\sqrt F=B-3+O(B^{-1})$ then gives Eq.~(\ref{m2tail}). The same expansion of $F$ and the relation $\lambda=B/(2c)$ give Eq.~(\ref{m2clambda}).

\section{Narrow-interval input for the small-moment tails}\label{app:smallm2}

We first derive the confinement fraction and action used for both bounds in Sec.~\ref{sec:m2small}, then calculate the second and higher even moments of the trial histories used for the upper bounds. Throughout this appendix, the limit $\ell\to0$ is taken at fixed $\mu$ before the large-$\mu$ limit. Our $\mu$ is the multiplier denoted by $\lambda$ in Ref.~\cite{BM2024interval}.

\subsection{Confinement fraction and action}\label{app:confinement}

Equation~(49) of Ref.~\cite{BM2024interval} determines the limiting energy fraction $\kappa$. Using the evenness of its integrand and extracting an elementary reference integral, we write it as

\begin{eqnarray} \label{smallm2integral}
2\pi\mu\kappa =2\int_0^\infty\frac{\ln(1+\mu^2\kappa k^2e^{-k^2})}{k^2}\,dk
&=&2\int_0^{\infty} \frac{\ln (1+\mu^2 \kappa k^2)}{k^2}\,dk + 2\int_0^\infty\frac{1}{k^2}
\ln\frac{1+\mu^2\kappa k^2e^{-k^2}}{1+\mu^2\kappa k^2}\,dk \nonumber \\
&=& 2\pi\mu\sqrt\kappa+2\int_0^\infty\frac{1}{k^2}
\ln\frac{1+\mu^2\kappa k^2e^{-k^2}}{1+\mu^2\kappa k^2}\,dk.
\end{eqnarray}

Let $K$ be the large root of $\mu^2\kappa K^2e^{-K^2}=1$, so that $K\simeq\sqrt{\ln(\mu^2\kappa)}$. Away from the small-$k$ region and the crossover near $K$, the remaining integrand is
\begin{equation}\label{smallm2integrals}
\frac{1}{k^2}\ln\frac{1+\mu^2\kappa k^2e^{-k^2}}{1+\mu^2\kappa k^2}
\simeq
\begin{cases}
-1,&(\mu\sqrt\kappa)^{-1}\ll k<K,\\[1mm]
-\ln(\mu^2\kappa k^2)/k^2,&k>K.
\end{cases}
\end{equation}
The first estimate follows by dropping $1$ in both logarithms; the second follows by dropping the exponentially small term in the numerator. The region $k=O((\mu\sqrt\kappa)^{-1})$ contributes $O((\mu\sqrt\kappa)^{-1})$. The crossover width is $O(K^{-1})$, since the logarithmic derivative of $\mu^2\kappa k^2e^{-k^2}$ at $K$ is $-2K+O(K^{-1})$. Both regions therefore contribute only $o(K)$. The plateau and tail in Eq.~(\ref{smallm2integrals}) each contribute $-2K$ to the correction in 
Eq.~(\ref{smallm2integral}):
\begin{equation}\label{smallm2plateautail}
-2K-2\int_K^\infty\frac{\ln(\mu^2\kappa k^2)}{k^2}\,dk+o(K)
=-2K-\frac{2[\ln(\mu^2\kappa K^2)+2]}{K}+o(K)
=-4K+o(K),
\end{equation}
where we have used the equality $\ln(\mu^2\kappa K^2)=K^2$. 
Consequently,
\begin{equation}\label{smallm2confinementasymptotic}
\mu(\sqrt\kappa-\kappa)\simeq\frac{2K}{\pi},\qquad
1-\kappa\simeq\frac{4\sqrt2}{\pi}\frac{\sqrt{\ln\mu}}{\mu},
\end{equation}
where the second relation uses $\kappa\to1$ and $\sqrt\kappa-\kappa\simeq(1-\kappa)/2$. This proves the first relation in Eq.~(\ref{smallm2interval}).

Let $s_{\mathrm{BM}}$ denote the limiting interval action. 
The positive-$\kappa$ solution starts at $\mu=\sqrt{4\pi}$ with $\kappa=s_{\mathrm{BM}}=0$, as follows by expanding Eq.~(\ref{smallm2integral}) at small $\kappa$. Integration by parts gives the exact quadrature
\begin{equation}\label{smallm2actionexact}
s_{\mathrm{BM}}(\mu)=\mu\kappa(\mu)-\int_{\sqrt{4\pi}}^\mu\kappa(z)\,dz
=-\mu[1-\kappa(\mu)]+\sqrt{4\pi}
+\int_{\sqrt{4\pi}}^\mu[1-\kappa(z)]\,dz.
\end{equation}
The boundary term is only $O(\sqrt{\ln\mu})$, whereas the integral yields
\begin{equation}\label{smallm2actionasymptotic}
s_{\mathrm{BM}}(\mu)\simeq\frac{4\sqrt2}{\pi}\int^\mu\frac{\sqrt{\ln z}}{z}\,dz
\simeq\frac{8\sqrt2}{3\pi}(\ln\mu)^{3/2},
\end{equation}
proving the second relation in Eq.~(\ref{smallm2interval}). The two coefficients $O(1)$ correct those in Eq.~(50) of Ref.~\cite{BM2024interval}. For a small finite interval, the required condition is $\ell\sqrt{\ln\mu}\ll1$, equivalent to Eq.~(51) of that reference.

\subsection{Second moment of the trial history}\label{app:trialsecond}

Write the terminal density as $\kappa\delta(x)+u_{\mathrm{out}}(x)$, where $u_{\mathrm{out}}$ is symmetric. Its one-sided Laplace transform follows from Eqs.~(36), (42), and (43) of Ref.~\cite{BM2024interval}. Specifically, set $k=-ip$, $p>0$, in those equations and take $\ell\to0$. Equation~(36) becomes
\begin{equation}\label{smallm2transformrelation}
\mu\kappa-p\mu^2\kappa\widehat u_{\mathrm{out}}(p)=-Q_-(-ip),
\end{equation}
while Eqs.~(42) and (43) give
\begin{equation}\label{smallm2scatteringfactor}
1-pQ_-(-ip)=F(p),\qquad
F(p)=\exp\!\left[\frac{p}{\pi}\int_0^\infty
\frac{\ln(1+\mu^2\kappa q^2e^{-q^2})}{q^2+p^2}\,dq\right].
\end{equation}
Here $Q_-$ is the one-sided Fourier transform defined in Eq.~(30) of that reference. Eliminating it gives
\begin{equation}\label{smallm2transform}
\widehat u_{\mathrm{out}}(p)\equiv\int_0^\infty e^{-px}u_{\mathrm{out}}(x)\,dx
=\frac{1+\mu\kappa p-F(p)}{\mu^2\kappa p^2}.
\end{equation}
The core contributes no second moment, so $m_2^{\mathrm{BM}}=2\widehat u_{\mathrm{out}}''(0^+)$. Because of the denominator $p^2$, we must expand $F$ through $p^4$, at fixed $\mu$.

As in Sec.~\ref{app:confinement}, first extract the reference contribution, now using
\begin{equation}\label{smallm2referenceexponent}
\frac{p}{\pi}\int_0^\infty\frac{\ln(1+\mu^2\kappa q^2)}{q^2+p^2}\,dq
=\ln(1+\mu\sqrt\kappa\,p).
\end{equation}
Thus $F(p)=(1+\mu\sqrt\kappa\,p)e^{\Phi(p)}$, where
\begin{equation}\label{smallm2residualexponent}
\Phi(p)=\frac{p}{\pi}\int_0^\infty\frac{L(q)}{q^2+p^2}\,dq,
\qquad L(q)=\ln\frac{1+\mu^2\kappa q^2e^{-q^2}}{1+\mu^2\kappa q^2}.
\end{equation}
Equation~(\ref{smallm2integral}) fixes its linear coefficient:
\begin{equation}\label{smallm2D}
\mathcal D\equiv-\frac{1}{\pi}\int_0^\infty\frac{L(q)}{q^2}\,dq
=\mu(\sqrt\kappa-\kappa)
\simeq\frac{2\sqrt2}{\pi}\sqrt{\ln\mu}.
\end{equation}
To obtain the higher coefficients without introducing divergent integrals, subtract the small-$q$ contribution $-\mu^2\kappa q^4/(1+\mu^2\kappa q^2)$ and define
\begin{equation}\label{smallm2R}
\mathcal R=\frac{1}{\pi}\int_0^\infty\frac{dq}{q^4}
\left[L(q)+\frac{\mu^2\kappa q^4}{1+\mu^2\kappa q^2}\right].
\end{equation}
Using $1/(q^2+p^2)=q^{-2}-p^2/[q^2(q^2+p^2)]$ and evaluating the subtracted rational integral gives
\begin{equation}\label{smallm2exponentexpansion}
\begin{aligned}
\Phi(p)&=-\mathcal Dp+\frac{\mu\sqrt\kappa\,p^3}{2(1+\mu\sqrt\kappa\,p)}-\mathcal R p^3+O(p^5)\\
&=-\mathcal Dp+\left(\frac{\mu\sqrt\kappa}{2}-\mathcal R\right)p^3
-\frac{\mu^2\kappa}{2}p^4+O(p^5).
\end{aligned}
\end{equation}
The remainder estimate holds at fixed $\mu$: the bracket in Eq.~(\ref{smallm2R}) is $O(q^6)$ as $q\to0$. The coefficient of $p^4$ in $(1+\mu\sqrt\kappa\,p)e^{\Phi(p)}$ is therefore
\begin{equation}\label{smallm2fourthcoefficient}
\frac{\mathcal D^4}{24}+\mathcal D\mathcal R
-\mu\sqrt\kappa\left(\frac{\mathcal D^3}{6}+\frac{\mathcal D}{2}+\mathcal R\right).
\end{equation}
Multiplying it by $-4/(\mu^2\kappa)$ yields the exact second moment in the narrow-interval limit:
\begin{equation}\label{smallm2momentexp}
m_2^{\mathrm{BM}}(\mu)=\frac{2\mathcal D^3/3+2\mathcal D+4\mathcal R}{\mu\sqrt\kappa}
-\frac{\mathcal D^4/6+4\mathcal D\mathcal R}{\mu^2\kappa}.
\end{equation}

To identify the leading term, note that the bracket in Eq.~(\ref{smallm2R}) lies between zero and $\min(q^4/8,q^2)$. Indeed, for fixed $z=\mu^2\kappa q^2$, the function $\ln[(1+ze^{-t})/(1+z)]+zt/(1+z)$ and its first derivative vanish at $t=0$, while its second derivative is between zero and $1/4$. Setting $t=q^2$ proves the $q^4/8$ bound; the $q^2$ bound follows because the logarithm is nonpositive. Hence $0\leq\mathcal R\leq1/(\sqrt2\pi)$, independently of $\mu$. Since $\mathcal D\to\infty$ but $\mathcal D/\mu\to0$, Eq.~(\ref{smallm2momentexp}) reduces to
\begin{equation}\label{smallm2secondasymptotic}
m_2^{\mathrm{BM}}(\mu)\simeq\frac{2\mathcal D^3}{3\mu}
\simeq\frac{32\sqrt2}{3\pi^3}\frac{(\ln\mu)^{3/2}}{\mu},
\end{equation}
which is Eq.~(\ref{smallm2trialmoment}). Equations~(\ref{smallm2integral}), (\ref{smallm2actionexact}), and (\ref{smallm2momentexp}), evaluated without a large-$\mu$ expansion, also specify the unexpanded trial-history calculation quoted in Sec.~\ref{sec:numn2}.

\subsection{Higher even moments}\label{app:trialhigher}

For a fixed even $n\geq2$, the numerator of Eq.~(\ref{smallm2transform}) must be expanded through $p^{n+2}$. A direct expansion of $\Phi$ produces coefficients growing as powers of $\mu$, which cancel after multiplication by $1+\mu\sqrt\kappa\,p$, as already seen at order $p^4$. The following alternative factorization makes these cancellations explicit at every fixed order.

Let $\rho>0$ solve
\begin{equation}\label{smallmnrho}
\mu^2\kappa\rho^2e^{\rho^2}=1,\qquad
\rho^{-1}=\mu\sqrt\kappa+O(\mu^{-1}).
\end{equation}
The numerator $1+\mu^2\kappa q^2e^{-q^2}$ has zeros at $q=\pm i\rho$. Extracting these zeros instead of those of the previous reference factor gives the exact representation
\begin{equation}\label{smallmnfactorization}
F(p)=(1+p/\rho)\exp\!\left[\frac{p}{\pi}\int_0^\infty
\frac{\widetilde L(q)}{q^2+p^2}\,dq\right],\qquad
\widetilde L(q)=\ln\frac{1+\mu^2\kappa q^2e^{-q^2}}{1+q^2/\rho^2}.
\end{equation}
The linear coefficient in this exponent is
\begin{equation}\label{smallmnlinearcoefficient}
\frac{1}{\pi}\int_0^\infty\frac{\widetilde L(q)}{q^2}\,dq
=-\mathcal D-\rho^{-1}+\mu\sqrt\kappa
=-\mathcal D+O(\mu^{-1}).
\end{equation}
All higher Taylor coefficients of the exponent remain bounded as $\mu\to\infty$, at each fixed order. To see why, the defining equation for $\rho$ gives
\begin{equation}\label{smallmnregularlog}
\widetilde L(q)=-q^2-\rho^2+
\ln\left[1+\rho^2\int_0^1 e^{t(q^2+\rho^2)}\,dt\right].
\end{equation}
This expression has uniformly bounded Taylor coefficients near $q=0$, with the nearby zeros removed. For real $q\geq1$, $|\widetilde L(q)|\leq q^2+\rho^2$; the integrals determining the cubic and higher odd coefficients then have convergent tails, bounded independently of $\mu$. Subtracting the small-$q$ Taylor terms before expanding the kernel, as in Sec.~\ref{app:trialsecond}, establishes the stated boundedness. The even coefficients are fixed by the same small-$q$ expansion.

It follows that the factor remaining after $1+p/\rho$ has, at every fixed order $j$, Taylor coefficient $(-\mathcal D)^j/j!\,[1+o(1)]$: terms containing a bounded quadratic or higher coefficient carry at least two fewer powers of $\mathcal D$. The coefficient of $p^{n+2}$ in $F(p)$ is consequently dominated by
\begin{equation}\label{smallmnhighercoefficient}
\frac{1}{\rho}\frac{(-\mathcal D)^{n+1}}{(n+1)!}[1+o(1)],
\end{equation}
since the term without $1/\rho$ is smaller by $O(\rho\mathcal D)$. Substitution into Eq.~(\ref{smallm2transform}) gives
\begin{equation}\label{smallmnmomentexp}
m_n^{\mathrm{BM}}(\mu)
=2(-1)^n\widehat u_{\mathrm{out}}^{(n)}(0^+)
\simeq\frac{2\mathcal D^{n+1}}{(n+1)\mu^2\kappa\rho}
\simeq\frac{2\mathcal D^{n+1}}{(n+1)\mu\sqrt\kappa},
\qquad \mu\to\infty,\quad n\ \text{fixed and even}.
\end{equation}
Using Eq.~(\ref{smallm2D}) and $\kappa\to1$ proves Eq.~(\ref{smallmntrialmoment}). In particular, $n=2$ recovers the leading term of Eq.~(\ref{smallm2momentexp}). The expansion at $p=0$ is performed before the large-$\mu$ limit; the result therefore follows from the transform and does not rely on inferring moment convergence from the pointwise plateau profile. Combining it with the action from Sec.~\ref{app:confinement} gives the upper-bound asymptotic used in Sec.~\ref{sec:higher}, where the assumptions needed to return to finite $\ell$ are stated.


\begin{thebibliography}{99}

\bibitem{BulinskiShashkin2007} A. Bulinski and A. Shashkin, \textit{Limit Theorems for Associated Random Fields and Related Systems} (World Scientific, Singapore, 2007).

\bibitem{GiraitisKoulSurgailis2012} L. Giraitis, H. L. Koul, and D. Surgailis, \textit{Large Sample Inference for Long Memory Processes} (Imperial College Press, London, 2012).

\bibitem{MerlevedePeligradUtev2019} F. Merlev\`ede, M. Peligrad, and S. Utev, \textit{Functional Gaussian Approximation for Dependent Structures} (Oxford University Press, Oxford, 2019).

\bibitem{CluselBertin2008} M. Clusel and E. Bertin, \href{https://doi.org/10.1142/S021797920804853X}{Int. J. Mod. Phys. B \textbf{22}, 3311 (2008)}.

\bibitem{Johansson1998} K. Johansson, \href{https://doi.org/10.1215/S0012-7094-98-09108-6}{Duke Math. J. \textbf{91}, 151 (1998)}.

\bibitem{LytovaPastur2009} A. Lytova and L. Pastur, \href{https://doi.org/10.1214/09-AOP452}{Ann. Probab. \textbf{37}, 1778 (2009)}.

\bibitem{CundenVivo2014} F. D. Cunden and P. Vivo, \href{https://doi.org/10.1103/PhysRevLett.113.070202}{Phys. Rev. Lett. \textbf{113}, 070202 (2014)}.

\bibitem{Forrester2023} P. J. Forrester, \href{https://doi.org/10.1214/23-PS15}{Probab. Surv. \textbf{20}, 170 (2023)}.

\bibitem{CundenFacchiVivo2016} F. D. Cunden, P. Facchi, and P. Vivo, \href{https://doi.org/10.1088/1751-8113/49/13/135202}{J. Phys. A: Math. Theor. \textbf{49}, 135202 (2016)}.

\bibitem{Valov2024} A. Valov, B. Meerson, and P. V. Sasorov, \href{https://doi.org/10.1088/1751-8121/ad1e1a}{J. Phys. A: Math. Theor. \textbf{57}, 065001 (2024)}.

\bibitem{Dean2019} D. S. Dean, P. Le Doussal, S. N. Majumdar, and G. Schehr, \href{https://doi.org/10.1088/1751-8121/ab098d}{J. Phys. A: Math. Theor. \textbf{52}, 144006 (2019)}.

\bibitem{Smith2021fermions} N. R. Smith, P. Le Doussal, S. N. Majumdar, and G. Schehr, \href{https://doi.org/10.21468/SciPostPhys.11.6.110}{SciPost Phys. \textbf{11}, 110 (2021)}.

\bibitem{LeDoussalSchehr2025} P. Le Doussal and G. Schehr, \href{https://doi.org/10.1007/s10955-025-03429-6}{J. Stat. Phys. \textbf{192}, 47 (2025)}.

\bibitem{Jona} L. Bertini, A. De Sole, D. Gabrielli, G. Jona-Lasinio, and C. Landim, \href{https://doi.org/10.1103/RevModPhys.87.593}{Rev. Mod. Phys. \textbf{87}, 593 (2015)}.

\bibitem{KMP1982} C. Kipnis, C. Marchioro, and E. Presutti, \href{https://doi.org/10.1007/BF01011740}{J. Stat. Phys. \textbf{27}, 65 (1982)}.

\bibitem{Spohn} H. Spohn, \textit{Large Scale Dynamics of Interacting Particles} (Springer-Verlag, Berlin, 1991).

\bibitem{DG2009a} B. Derrida and A. Gerschenfeld, \href{https://doi.org/10.1007/s10955-009-9772-7}{J. Stat. Phys. \textbf{136}, 1 (2009)}.

\bibitem{DG2009b} B. Derrida and A. Gerschenfeld, \href{https://doi.org/10.1007/s10955-009-9830-1}{J. Stat. Phys. \textbf{137}, 978 (2009)}.

\bibitem{KrMe} P. L. Krapivsky and B. Meerson, \href{https://doi.org/10.1103/PhysRevE.86.031106}{Phys. Rev. E \textbf{86}, 031106 (2012)}.

\bibitem{MS2013} B. Meerson and P. V. Sasorov, \href{https://doi.org/10.1088/1742-5468/2013/12/P12011}{J. Stat. Mech. (2013) P12011}.

\bibitem{MS2014} B. Meerson and P. V. Sasorov, \href{https://doi.org/10.1103/PhysRevE.89.010101}{Phys. Rev. E \textbf{89}, 010101(R) (2014)}.
    
\bibitem{VMS2014} A. Vilenkin, B. Meerson, and P. V. Sasorov, \href{https://doi.org/10.1088/1742-5468/2014/06/P06007}{J. Stat. Mech. (2014) P06007}.

\bibitem{Novikov} S. P. Novikov, S. V. Manakov, L. P. Pitaevskii, and V. E. Zakharov, \textit{Theory of Solitons: The Inverse Scattering Method} (Consultants Bureau, New York, 1984).

\bibitem{BSM2022a} E. Bettelheim, N. R. Smith, and B. Meerson, \href{https://doi.org/10.1103/PhysRevLett.128.130602}{Phys. Rev. Lett. \textbf{128}, 130602 (2022)}.

\bibitem{BSM2022b} E. Bettelheim, N. R. Smith, and B. Meerson, \href{https://doi.org/10.1088/1742-5468/ac8a4d}{J. Stat. Mech. (2022) 093103}.

\bibitem{MMS2022} K. Mallick, H. Moriya, and T. Sasamoto, \href{https://doi.org/10.1103/PhysRevLett.129.040601}{Phys. Rev. Lett. \textbf{129}, 040601 (2022)}.

\bibitem{Grabsch2022} A. Grabsch, A. Poncet, P. Rizkallah, P. Illien, and O. B\'enichou, \href{https://doi.org/10.1126/sciadv.abm5043}{Sci. Adv. \textbf{8}, eabm5043 (2022)}.

\bibitem{BM2024} E. Bettelheim and B. Meerson, \href{https://doi.org/10.1103/PhysRevE.110.014101}{Phys. Rev. E \textbf{110}, 014101 (2024)}.

\bibitem{BM2024interval} E. Bettelheim and B. Meerson, \href{https://doi.org/10.1088/1742-5468/ad8b39}{J. Stat. Mech. (2024) 113204}.

\bibitem{MMS2024} K. Mallick, H. Moriya, and T. Sasamoto, \href{https://doi.org/10.1088/1742-5468/ad485e}{J. Stat. Mech. (2024) 074001}.

\bibitem{CS2001} A. I. Chernykh and M. G. Stepanov, \href{https://doi.org/10.1103/PhysRevE.64.026306}{Phys. Rev. E \textbf{64}, 026306 (2001)}.

\bibitem{GVE2019} T. Grafke and E. Vanden-Eijnden, \href{https://doi.org/10.1063/1.5084025}{Chaos \textbf{29}, 063118 (2019)}.

\bibitem{SGMG2022} T. Schorlepp, T. Grafke, S. May, and R. Grauer, \href{https://doi.org/10.1098/rsta.2021.0051}{Phil. Trans. R. Soc. A \textbf{380}, 20210051 (2022)}.

\bibitem{STGS2023} T. Schorlepp, S. Tong, T. Grafke, and G. Stadler, \href{https://doi.org/10.1007/s11222-023-10307-2}{Stat. Comput. \textbf{33}, 137 (2023)}.

\bibitem{SSM2023} T. Schorlepp, P. Sasorov, and B. Meerson, \href{https://doi.org/10.1088/1742-5468/ad0a94}{J. Stat. Mech. (2023) 123202}.

\bibitem{BSG2024} S. Burekovi{\'c}, T. Sch{\"a}fer, and R. Grauer, \href{https://doi.org/10.1103/PhysRevLett.133.077202}{Phys. Rev. Lett. \textbf{133}, 077202 (2024)}.

\bibitem{DeGiuli25} E. De Giuli, \href{https://doi.org/10.1103/vlx8-spc6}{Phys. Rev. E \textbf{112}, 024142 (2025)}.

\bibitem{SG2026} T. Schorlepp and T. Grafke, \href{https://doi.org/10.1137/25M1743867}{SIAM/ASA J. Uncertain. Quantif. \textbf{14}, 920 (2026)}.

\bibitem{BKP2012} G. Bunin, Y. Kafri, and D. Podolsky, \href{https://doi.org/10.1209/0295-5075/99/20002}{EPL \textbf{99}, 20002 (2012)}.

\bibitem{Tailleur} J. Tailleur, J. Kurchan, and V. Lecomte, \href{https://doi.org/10.1103/PhysRevLett.99.150602}{Phys. Rev. Lett. \textbf{99}, 150602 (2007)}.

\bibitem{EK2004} V. Elgart and A. Kamenev, \href{https://doi.org/10.1103/PhysRevE.70.041106}{Phys. Rev. E \textbf{70}, 041106 (2004)}.

\bibitem{GGS2015} T. Grafke, R. Grauer, and T. Sch\"afer, \href{https://doi.org/10.1088/1751-8113/48/33/333001}{J. Phys. A: Math. Theor. \textbf{48}, 333001 (2015)}.

\bibitem{MKV2016} B. Meerson, E. Katzav, and A. Vilenkin, \href{https://doi.org/10.1103/PhysRevLett.116.070601}{Phys. Rev. Lett. \textbf{116}, 070601 (2016)}.

\bibitem{GBPC2017} L. S. Grigorio, F. Bouchet, R. M. Pereira, and L. Chevillard, \href{https://doi.org/10.1088/1751-8121/aa51a3}{J. Phys. A: Math. Theor. \textbf{50}, 055501 (2017)}.

\bibitem{AMPV2022} G. B. Apolin\'ario, L. Moriconi, R. M. Pereira, and V. J. Valad\~ao, \href{https://doi.org/10.1016/j.physleta.2022.128360}{Phys. Lett. A \textbf{449}, 128360 (2022)}.

\bibitem{GB2023} G. T. Goedert and L. Biferale, \href{https://doi.org/10.1140/epje/s10189-023-00392-6}{Eur. Phys. J. E \textbf{46}, 132 (2023)}.

\bibitem{Plessix} R.-E. Plessix, \href{https://doi.org/10.1111/j.1365-246X.2006.02978.x}{Geophys. J. Int. \textbf{167}, 495 (2006)}.

\bibitem{LN} D. C. Liu and J. Nocedal, \href{https://doi.org/10.1007/BF01589116}{Math. Program. \textbf{45}, 503 (1989)}.

\bibitem{NW} J. Nocedal and S. J. Wright, \textit{Numerical Optimization}, 2nd ed. (Springer, New York, 2006).
    
\bibitem{Hestenes} M. R. Hestenes, \href{https://doi.org/10.1007/BF00927673}{J. Optim. Theory Appl. \textbf{4}, 303 (1969)}.

\bibitem{Powell} M. J. D. Powell, in \textit{Optimization}, edited by R. Fletcher (Academic Press, London, 1969), p.~283.






\bibitem{nonconvexnumerics} This does not preclude stationary solutions of the MFT boundary-value problem on the nonconvex branch, and a suitable reparametrization of the observable can remove the obstruction in a tilted formulation~\cite{AG2021}, but the reparametrization must be chosen with care.


\bibitem{AG2021} M. Alqahtani and T. Grafke, \href{https://doi.org/10.1088/1751-8121/abe67b}{J. Phys. A: Math. Theor. \textbf{54}, 175001 (2021)}.

\bibitem{JAX} J. Bradbury, R. Frostig, P. Hawkins, M. J. Johnson, Y. Katariya, C. Leary, D. Maclaurin, G. Necula, A. Paszke, J. VanderPlas, S. Wanderman-Milne, and Q. Zhang, JAX: composable transformations of Python+NumPy programs, version 0.3.13, \url{https://github.com/jax-ml/jax} (2018).
\bibitem{SG1969} D. L. Scharfetter and H. K. Gummel, \href{https://doi.org/10.1109/T-ED.1969.16566}{IEEE Trans. Electron Devices \textbf{16}, 64 (1969)}.

\bibitem{Patankar} S. V. Patankar, \textit{Numerical Heat Transfer and Fluid Flow} (Hemisphere, Washington, DC, 1980).

\bibitem{Heydecker2026} D. Heydecker, 
\href{https://arxiv.org/abs/2605.26652}{arXiv:2605.26652 (2026)}.


\end{thebibliography}
\end{document}